\documentclass[12pt]{article}
\usepackage[latin9]{inputenc}
\usepackage{mathtools}
\usepackage{algorithm2e}
\usepackage{amsmath}
\usepackage{overcite}
\usepackage{graphicx}
\usepackage{mathbbol}
\usepackage{xcolor}
\usepackage{xfrac}

\newcommand{\W}{\mathcal{W}}

\usepackage{scicite}

\usepackage{times}

\newenvironment{sciabstract}{%
\begin{quote} \bf}
{\end{quote}}

\title{Network community detection and clustering with random walks}

\author
{Aditya Ballal,$^{1,2}$ Willow B. Kion-Crosby,$^{3}$ Alexandre V. Morozov$^{1,2,\ast}$\\
\\
\normalsize{$^1$ Department of Physics and Astronomy, Rutgers University, Piscataway, NJ 08854, USA}\\
\normalsize{$^2$ Center for Quantitative Biology, Rutgers University, Piscataway, NJ 08854, USA}\\
\normalsize{$^3$ Helmholtz Institute for RNA-based Infection Research, W{\"u}rzburg 97080, Germany}\\
\\
\normalsize{$^\ast$To whom correspondence should be addressed; E-mail: \texttt{morozov@physics.rutgers.edu}}
}

\date{}

\begin{document} 


\baselineskip24pt


\maketitle


\begin{sciabstract}
We present a novel approach to partitioning network nodes into non-overlapping communities -- a key step in revealing network modularity and hierarchical organization. Our methodology, applicable to networks with both weighted and unweighted symmetric edges, uses random walks to explore neighboring nodes in the same community. The walk-likelihood algorithm (WLA) produces an optimal partition of network nodes into a given number of communities. The walk-likelihood community finder (WLCF) employs WLA to predict both the optimal number of communities and the corresponding network partition. We have extensively benchmarked both algorithms, finding that they outperform or match other methods in terms of the modularity of predicted partitions and the number of links between communities. Making use of the computational efficiency of our approach, we investigated a large-scale map of roads and intersections in the state of Colorado. Our clustering yielded geographically sensible boundaries between neighboring communities.
\end{sciabstract}


\section*{Introduction}

Many complex systems in human society, science and technology can be represented by networks -- a set of $N$ vertices linked by edges~\cite{Albert2002,Barrat2008,Newman2010}. Examples include the Internet, the World Wide Web, transportation networks, food webs, social networks, and biochemical and genetic networks in biology. These complex networks often contain distinct groups,
with more edges between nodes within the same group than between nodes belonging to different groups.
Detecting such distinct groups of nodes, called network communities, has attracted 
considerable attention in the literature~\cite{Girvan_2002,Radicchi2004,Reichardt2004,Newman_2004,Newman_2004b,Palla2005,Walktrap2005,Newman_2006,Raghavan_2007,Blondel_2008}.
Parsing complex networks into communities provides useful information about the hierarchical structure of the network. For example, in gene co-expression networks communities represent gene modules, with genes in the same module acting together to carry out high-level biological functions such as stress response~\cite{Gasch2000}. Protein-protein interaction networks
are also characterized by pronounced modularity which may have been shaped by adaptive evolution~\cite{Mihalik2011}.
In the context of social networks, communities represent groups of people with similar interests and behavioral patterns.

Despite clear intuition behind the network community concept, mathematical definitions of network communities are somewhat elusive. A widely accepted quantitative definition of the community structure in a network is based on the modularity score~\cite{Newman_2004} (Methods).
The notion of the modularity score plays a key role in several algorithms for network community detection~\cite{Clauset_2004,Newman_2006,Raghavan_2007,Blondel_2008}.
Commonly used network community detection methods include Edge Betweenness~\cite{Girvan_2002}, Fastgreedy~\cite{Clauset_2004}, Infomap~\cite{Rosvall2007}, Label
Propagation~\cite{Raghavan_2007}, Leading Eigenvector~\cite{Newman_2006}, Multilevel~\cite{Blondel_2008}, Spinglass~\cite{Reichardt2004}, and Walktrap~\cite{Walktrap2005}.
These methods were benchmarked for computational efficiency and prediction accuracy by Yang et al. using an extensive set of artificially generated networks~\cite{Yang2016}.
Besides the modularity score, we employ two additional measures used to investigate network partitioning into clusters: the internal edge density and the cut ratio~\cite{Leskovec2010,Yang2013} (Methods).

Network community detection is conceptually similar to clustering and data dimensionality reduction, which have a long history of development in machine learning and artificial intelligence communities~\cite{Bishop:2006}. Some of the state-of-the-art approaches for data clustering and visualization are rooted in the ideas borrowed from random walks and diffusion theory.
Specifically, non-negative matrix factorization (NMF) is a powerful clustering method, originally developed to provide decompositions into interpretable features in
visual recognition and text analysis tasks~\cite{Lee1999,Lee2001,nndsvd,NMF4}. NMF is based on decomposing a non-negative matrix $X_{N_1 \times N_2}$ into two non-negative matrices $L_{N_1 \times m}$ and $R_{m \times N_2}$: $X=LR$.
To cluster a graph into $m$ communities using NMF, the adjacency matrix of the graph $X$ is factorized into $L$ and $R$, and each node is assigned to the community with the largest
matrix element in the corresponding row of $L$ ($N_1 = N_2 = N$ in this case).
An algorithm closely related to NMF and based on analyzing the eigenvalues and eigenvectors of the graph Laplacian is called spectral clustering~\cite{NMF3,Luxburg2007}.
Finally, we note a dimensionality reduction technique based on diffusion maps, which uses random walks to project datapoints
into a lower-dimensional space~\cite{Coifman2005,Coifman2006,Delaporte2008}.

Here we propose two novel methods for clustering and network community detection. The first method, which we call the walk-likelihood algorithm (WLA), leverages information provided by
random walks to produce a partition of datapoints or network nodes into $m$ non-overlapping communities, where the number of communities is known \textit{a priori}.
Unlike previous algorithms that employ random walks and diffusion (either explicitly or implicitly, through spectral decomposition of the graph Laplacian) in network community detection~\cite{Walktrap2005,Newman_2006,Rosvall2007}, dimensionality reduction~\cite{Coifman2005,Coifman2006,Delaporte2008}, and spectral clustering~\cite{NMF3,Luxburg2007},
our approach is based on Bayesian inference of network properties as the network is explored by random walks~\cite{Willow}. One of these properties is the
posterior probability for each node to belong to one of the $m$ network communities. Instead of relying on a finite sample of random walks, we sum over all random paths with a
given number of steps, producing network community assignments for each node that are free of sampling noise. WLA is used as the main ingredient in our second
algorithm, walk-likelihood community finder (WLCF), which predicts the optimal number of clusters (or network communities) $m_\text{opt}$
using global moves that involve bifurcation and merging of communities, and employs WLA to refine node community assignments at each step. We have subjected both WLA and WLCF to extensive
testing on artificial networks against several of the state-of-the-art algorithms mentioned above. After establishing its superior performance compared to the other algorithms, we have
applied WLCF to several real-world networks, including a large-scale network of roads and intersections in the Colorado state.

\section*{Results}

\subsection*{Walk-likelihood algorithm}

Let us consider a network with $N$ nodes labeled $n=1 \dots N$. Let $A_{N\times N}$ be the transition
matrix of the network, where $A_{n'n} = P(n \rightarrow n') $ is the probability
to jump from node $n$ to node $n'$ in a single step (see Methods for details). We define a matrix
$U_{N \times m}$ to partition the network into $m$ communities labeled by $c =1 \dots m$, such
that each element $U_{nc}=1$ if and only if $n \in c$, and $0$ otherwise. The weighted size of each community $c$ can then be computed as $\W_{c} =\sum_{n=1}^N w_n U_{nc}$,
where $w_n$ is the connectivity of node $n$ (Methods).
Next, we define a matrix $V_{N \times m}$ such that

\begin{equation} \label{eq:3}
V_{nc} = \sum_{l=1}^{l_{max}} \left(\sum_{n'=1}^N (A^{l})_{nn'} w_{n'} U_{n'c}\right).
\end{equation}
Note that $V_{nc}/\W_{c}$ is the expected number of times, \emph{per random walk}, that the node $n$ is visited by random walks with $l_{max}$ steps which start from nodes $n'$ in community $c$, where the nodes $n' \in c$ are chosen randomly with probability $P(n') = w_{n'}/\W_c$. Note that node $n$ does not have to be in the same community $c$ as nodes $n'$, although the expected number of visits to node $n$ should be higher if this is the case. Furthermore, the expected number of visits per random walk to node $n$ given by Eq.~\eqref{eq:3} corresponds to the number of visits that would be observed when the total number of random walks with $l_{max}$ steps that originate from community $c$, $G_c$, is very large: $G_c \to \infty$.  
Then the total number of visits to node $n$ is given by $\tilde{V}_{nc} = G_{c} V_{nc}/\W_c$. The community identity of node $n$ can be inferred probabilistically using Eq.~\eqref{eq:2} (Methods):
\begin{equation} \label{eq:4}
P(n \in c|\{ \tilde{V}_{nc'} \}_{c'=1}^m, \{ \ell_{c'c} \}_{c'=1}^m) = \frac{1}{\mathcal{Z}}  {\prod_{c'=1}^m \mathcal{P} \left(\tilde{V}_{nc'}, \frac{w_n \ell_{c'c}}{\W_{c}}\right)},
\end{equation}
where $\ell_{c'c} = (\tilde{V}^{T}U)_{c'c} = G_{c'}(V^{T}U)_{c'c}/\W_{c'}$ is the total number of steps in community $c$ of $G_{c'}$ random walks that originate in community $c'$
(equal to the total number of visits to nodes in community $c$),
and $\mathcal{Z}$ is the normalization constant. Omitting the conditional dependencies for simplicity, Eq.~\eqref{eq:4} can be rewritten as:
\begin{equation} \label{eq:logPnc}
\log P(n\in c) = \sum_{c'=1}^m \frac{G_{c'}}{\W_{c'}} \left( V_{nc'} \log Q_{c'c} - Q_{c'c} w_{n} \right) + H(n) - \log \mathcal{Z},
\end{equation}
where
\begin{equation} \label{eq:Q}
Q_{c'c} = \frac{\ell_{c'c} \W_{c'}}{G_{c'}\W_c} = \frac{(V^{T}U)_{c'c}}{\W_{c}}    
\end{equation}
and
\begin{equation}
H(n) = \sum_{c'=1}^m \left[ \frac{G_{c'}}{\W_{c'}} \log \left(\frac{G_{c'}}{\W_{c'}}\right) - \log \left(\left(\frac{G_{c'}V_{nc'}}{\W_{c'}}\right)! \right) \right]
\end{equation}
is independent of the community index. 

We find it convenient to parameterize $G_c$ as $G_c = s g_c$, with $s \rightarrow \infty$ and finite relative weights $g_c$ (the choice of $g_c$ is discussed below). 
Then Eq.~\eqref{eq:4} can be written as
\begin{equation} \label{eq:Pnc}
P(n \in c) = \lim_{s \rightarrow \infty} \frac{e^{s F_{nc}}}{\sum_{c'=1}^m e^{s F_{nc'}}},
\end{equation}
where $F_{nc}$ is given by Eq.~\eqref{eq:logPnc}:
\begin{equation} \label{eq:5}
F_{nc} = \sum_{c'=1}^m \frac{g_{c'}}{\W_{c'}} \left[ V_{nc'} \log Q_{c'c} - Q_{c'c}w_{n} \right].
\end{equation}
In the $s \rightarrow \infty$ limit, the sum in the denominator of Eq.~\eqref{eq:Pnc} is dominated by a single term with the largest $F_{nc'}$, so that
Eq.~\eqref{eq:Pnc} simplifies to

\begin{equation} \label{eq:6}
P(n \in c) = \delta_{c \tilde{c}} \text{ for } \tilde{c}=  \text{argmax}_{c''} F_{nc''}.
\end{equation}
Equation~\eqref{eq:6} allows us to reassign community identities for each node $n$. These community identities are then used to construct the updated matrix $U$ for the next iteration
of the algorithm.

\noindent
\\ \textbf{Choice of }$g_{c}$\textbf{. }The relative weights $g_{c}$ determine the fraction of random walks that start from community $c$.
To remove community-dependent sampling biases, we set $g_{c}$ so that the mean number of visits to node $n \in c$
from all random walks starting in the community $c$ is independent of its
parameters. Note that according to Eq.~\eqref{eq:mean}, the mean number of visits to
a node $n \in c$ is $\ell_{cc}w_{n}/\W_c = s g_{c}Q_{cc}w_{n}/\W_c$. Thus, setting $g_{c}=\W_c/Q_{cc}$ ensures that the mean number of visits is $s w_{n}$,
which is independent of the community index $c$ and depends only on the connectivity of node $n$.


\noindent
\\ \textbf{Convergence Criterion}\textbf{. }To determine how similar the updated assignment of nodes to communities is to the previous one, we use the normalized mutual information (NMI)~\cite{NMI}
between the current partition $U$ and the previous partition $U'$ (Eq.~\eqref{eq:nmi} in Methods).
We terminate the iterative node reassignment process if the NMI between partitions obtained in subsequent iterations is greater than $0.99$.

\noindent
\\ The iterative node reassignment procedure can be summarized as follows:
\vspace{20pt}

\hrule
\hrule
\hrule
\vspace{5pt}
\noindent
\textbf{WALK-LIKELIHOOD ALGORITHM}
\vspace{5pt}
\hrule
\hrule
\hrule
\vspace{5pt}
\noindent
\textbf{INPUT:} \\
Network with $N$ nodes\\
$A_{N\times N}$: Transition matrix of the network\\
$w_n$: Connectivity of each node $n = 1 \dots N$\\
$U'_{N\times m}$: Initial guess of the partition of the network into $m$ communities
\vspace{5pt}
\hrule
\vspace{5pt}
\noindent
\textbf{do:}
\begin{enumerate}
\item $V_{nc} \leftarrow \sum_{l=1}^{l_{max}} \left(\sum_{n'=1}^N (A^{l})_{nn'} w_{n'} {U'}_{n'c} \right)$ \hfill
Eq.~\eqref{eq:3}
\item $Q_{cc'} \leftarrow (V^{T} U')_{cc'} / \sum_{n=1}^N w_{n} {U'}_{nc'}$ \hfill
Eq.~\eqref{eq:Q}
\item $F_{nc} \leftarrow \sum_{c'=1}^m Q_{c'c'}^{-1} \left[ V_{nc'} \log Q_{c'c} - Q_{c'c}w_{n} \right]$ \hfill Eq.~\eqref{eq:5} with $g_c = \W_c / Q_{cc}$
\item $U_{nc} \leftarrow \delta_{\tilde{c}_{n}c}$ for $\tilde{c}_{n} = \text{argmax}_{c''} F_{nc''}$ \hfill Eq.~\eqref{eq:6}
\item Compute $\text{NMI}(U,U')$ \hfill Eq.~\eqref{eq:nmi}
\item $U' \leftarrow U$
\end{enumerate}
\noindent
\textbf{while not converged} [$\text{NMI}(U,U') \le 0.99$]
\vspace{5pt}
\hrule
\vspace{5pt}
\noindent
\textbf{OUTPUT:} $U_{N\times m}$: Optimal partition of the network into $m$ communities
\vspace{5pt}
\hrule

\subsection*{Walk-likelihood community finder}

\begin{figure}
\centering
\includegraphics[scale=0.15]{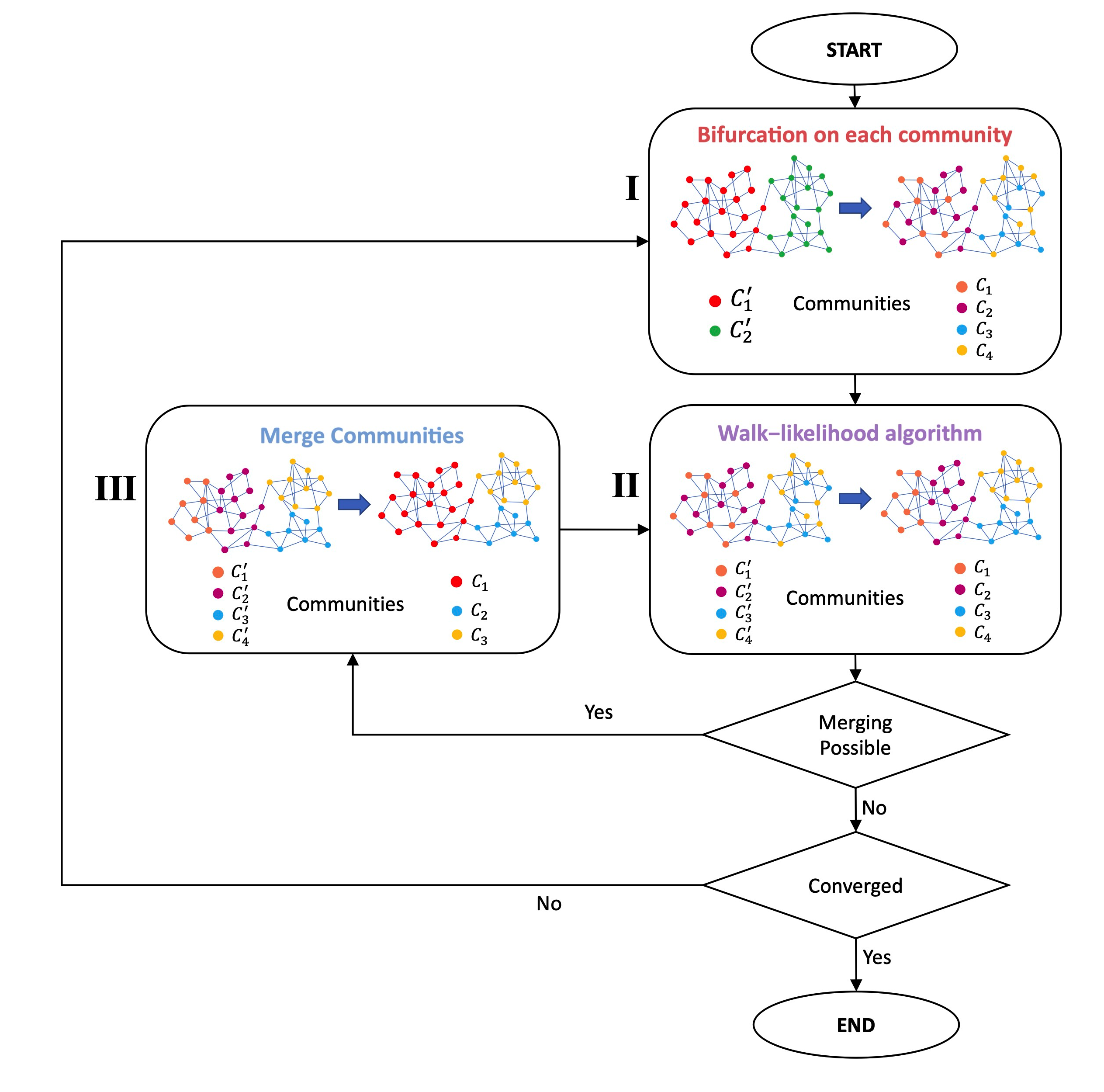}
\caption{\textbf{The flowchart of the WLCF algorithm.} The key steps of the algorithm include: random community bifurcation in the beginning of the outer loop iteration (panel I); application of the walk-likelihood algorithm (panel II); merging communities on the basis of the changes in the modularity score (panel III).}
\label{fig:flow}
\end{figure}

Using the walk-likelihood algorithm (WLA) described above, we have developed the walk-likelihood community finder (WLCF) -- an algorithm for partitioning a network into communities when the number of communities is not known \textit{a priori}. We initialize the WLCF algorithm by assuming that the whole network is a single community. The flowchart of the algorithm is shown in Fig.~\ref{fig:flow}, with each major step explained in detail below:

\vspace{5pt}
\noindent \textbf{Outer loop:}
\begin{itemize}
\item [\textbf{I.}] \textbf{Bifurcation:} We bifurcate each network community randomly into two communities. This is illustrated in Fig.~\ref{fig:flow}, panel I, where community $C'_1$ bifurcates into communities $C_1$ and $C_2$, and community $C'_2$ bifurcates into communities $C_3$ and $C_4$. Note that this step bifurcates the network into two communities at the start of the algorithm.
\item [\textbf{II.}] \textbf{Inner Loop:} The inner loop consists of three consecutive steps. The loop is terminated if step 2 conditions are not met.
\begin{itemize}
    \item[\textbf{1.}] \textbf{Walk-likelihood algorithm:} The walk-likelihood algorithm is run to obtain a more accurate partition of the network (Fig.~\ref{fig:flow}, panel II). Note that the number of communities $m$ does not change in this step.
    \item[\textbf{2.}] \textbf{Criteria for merging communities:} To check if the current division of the network into $m$ communities is optimal, we compute modularity scores~\cite{Newman_2004}
    for all $m$ communities. Then, for $m \choose 2$ pairs of communities, we check if combining any pair of communities $c$ and $c'$ increases the modularity score of the partition.
    The change in the modularity score after merging communities $c$ and $c'$ is given by 
    \begin{equation}
        \Delta M_{cc'} = 2 (e_{cc'} - a_c a_{c'}),
    \end{equation}
    where $e_{cc'} = (U^T \tilde{A} U)_{cc'}/\sum_{n=1}^N w_n$ and $a_c = \sum_{n=1}^N U_{nc}w_n/\sum_{n=1}^N w_n$ ($\tilde{A}$ is the symmetric adjacency matrix with 1 denoting edges
    and 0 everywhere else). Note that these definitions generalize the modularity score (Eq.~\eqref{M:score} in Methods) to networks with weighted edges. 
    If there exists at least one pair of communities such that $\Delta M_{cc'}>0$, we proceed to step 3 of the inner loop where one pair of communities is merged, otherwise we exit the inner loop.
    \item[\textbf{3.}] \textbf{Merging Communities:} If step 2 conditions are met, we merge the pair of communities $c_1$ and $c_2$ with the largest increase in the modularity score $M_{c_1 c_2}$.
    This is illustrated in Fig.~\ref{fig:flow}, panel III, where communities $C'_1$ and $C'_2$ merge to form $C_1$. 
\end{itemize}
\item [\textbf{III.}] \textbf{Convergence Criteria:} The outer loop is terminated if the number of communities in the partitions obtained in subsequent iterations of the outer loop remains constant and the NMI between the communities in the current and the previous partitions is greater than $0.99$.
The algorithm also stops if the modularity score of the partition decreases by more than $0.01$ in subsequent iterations, or if the maximum number of iterations has been reached.
\end{itemize}

\noindent \textbf{Elimination of spurious bifurcation-merge cycles.}
The WLCF algorithm can get into a loop where a community $c$ is bifurcated into $c_1$ and $c_2$ in step I and then $c_1$ and $c_2$ merge again in step 3 of the inner loop (step II of the outer loop) to form the same community $c$. This indicates that community $c$ cannot be bifurcated any further. In order to avoid such bifurcation-merge cycles, we check if there are any matches between the communities in the current partition and those in the previous partition, by calculating the following score:
\begin{equation}
    E_{cc'} = 1 - \frac{2\sum_{i=1}^N U_{ic}U'_{ic'}}{\sum_{i=1}^N (U_{ic}+ U'_{ic'})}
\end{equation}
between the communities $c$ and $c'$ of the current partition ($U$) and the previous partition ($U'$), respectively. If $E_{cc'} < 0.01$, we assume that the communities $c$ and $c'$ are the same and conclude that further bifurcations of the community $c$ are not possible. Thus, all communities $c$ of the current partition for which there exists a corresponding community $c'$ in the previous partition such that $E_{cc'} < 0.01$, are not bifurcated in the subsequent iteration of the WLCF algorithm (step I of the outer loop).

\subsection*{Synthetic Networks}
To test the performance of WLA and WLCF algorithms in a controlled setting using realistic networks with tunable properties, we have generated a comprehensive set of Lancichinetti, Fortunato and Radicchi (LFR) benchmark graphs~\cite{Lancichinetti_2008}.
The LFR benchmark was specifically created to provide a challenging test for community detection algorithms. It was recently used to test many
state-of-the-art algorithms in a rigorous comparative analysis~\cite{Yang2016}. Similar to real-world networks, LFR networks are characterized by power-law distributions of the node degree and community size. Each node in a given LFR network has a fixed mixing parameter $\mu = {\sum_{i=1}^N k^\text{ext}_i}/{\sum_{i=1}^N k_i}$, where $k_i^\text{ext}$ is the number of links between node $i$ and nodes in all other communities and $k_i$ is the total number of links of node $i$. Thus, every node 
shares a fraction $1 - \mu$ of its links with the other nodes in its community and a fraction $\mu$ with the rest of the network~\cite{Lancichinetti_2008}.
Note that $\mu = 0$ corresponds to the communities that are completely isolated from one another, while $\mu < 1/2$ results in well-defined communities in which each node
has more connections with the nodes in its own community than with the rest of the graph. Generally speaking, network communities become more difficult to detect as $\mu$ increases.

The parameters of the networks in our LFR benchmark set are summarized in Table~S1. These parameters were chosen to enable direct comparisons with the large-scale evaluation of community detection algorithms carried out by Yang et al.~\cite{Yang2016}. In order to investigate algorithm performance on larger networks, we have also added graphs with $N=5 \times 10^4$ and $N=10^5$ to our implementation of the LFR benchmark.
For each value of $N$, we have created networks with $25$ different mixing parameters $\mu$ ranging from $0.03$ to $0.75$. For each value of $N$ and $\mu$,
$20$ independent network realizations were created for networks with $N=5 \times 10^4$ and $N=10^5$; for all smaller networks, $10^2$ independent network realizations were created.
We used the Github package \texttt{LFR-Benchmark\_UndirWeightOvp} by eXascale Infolab (\texttt{https://github.com/eXascaleInfolab}) 
to generate the LFR benchmark networks. 

First, we have used a single realization of the LFR network with $\mu=0.15$ and $N=10^3$ to study the effects of $l_{max}$ on the network exploration (Fig.~S1).
Similar to diffusion maps~\cite{Coifman2005,Coifman2006,Delaporte2008}, the value of $l_{max}$ is related to the scale of the network structures explored by random
walks: lower $l_{max}$ values create a bias towards local exploration, while higher $l_{max}$ values enable global exploration of the entire network and transitions between
communities. The natural upper cutoff for $l_{max}$ is the network diameter, which is often $\sim \log N$ in scale-free, real-world networks~\cite{Albert2002,Barrat2008,Newman2010}.
Indeed, we observe that small $l_{max}$ values lead to more visits to nodes in the same community as the starting node (compared to nodes in all other communities)
as local network neighborhoods are explored (Fig.~S1).
However, the exploration is noisy since many nodes cannot be reached by short random walks, even if they belong to the same community. As $l_{max}$ increases, the difference between
visiting nodes in the two categories decreases, but the uncertainties in the number of visits decrease at the same time. For very large $l_{max}$ values, the whole network is explored.
Overall, we conclude that either using an intermediate value of $l_{max}$ or alternating between an intermediate and a low value should lead to reasonable performance.
As with hyperparameter settings in many other algorithms, finding an acceptable range of $l_{max}$ values may require some numerical experimentation.

\begin{figure}
\centering
\hspace*{-1.0cm}
\includegraphics[scale=0.5]{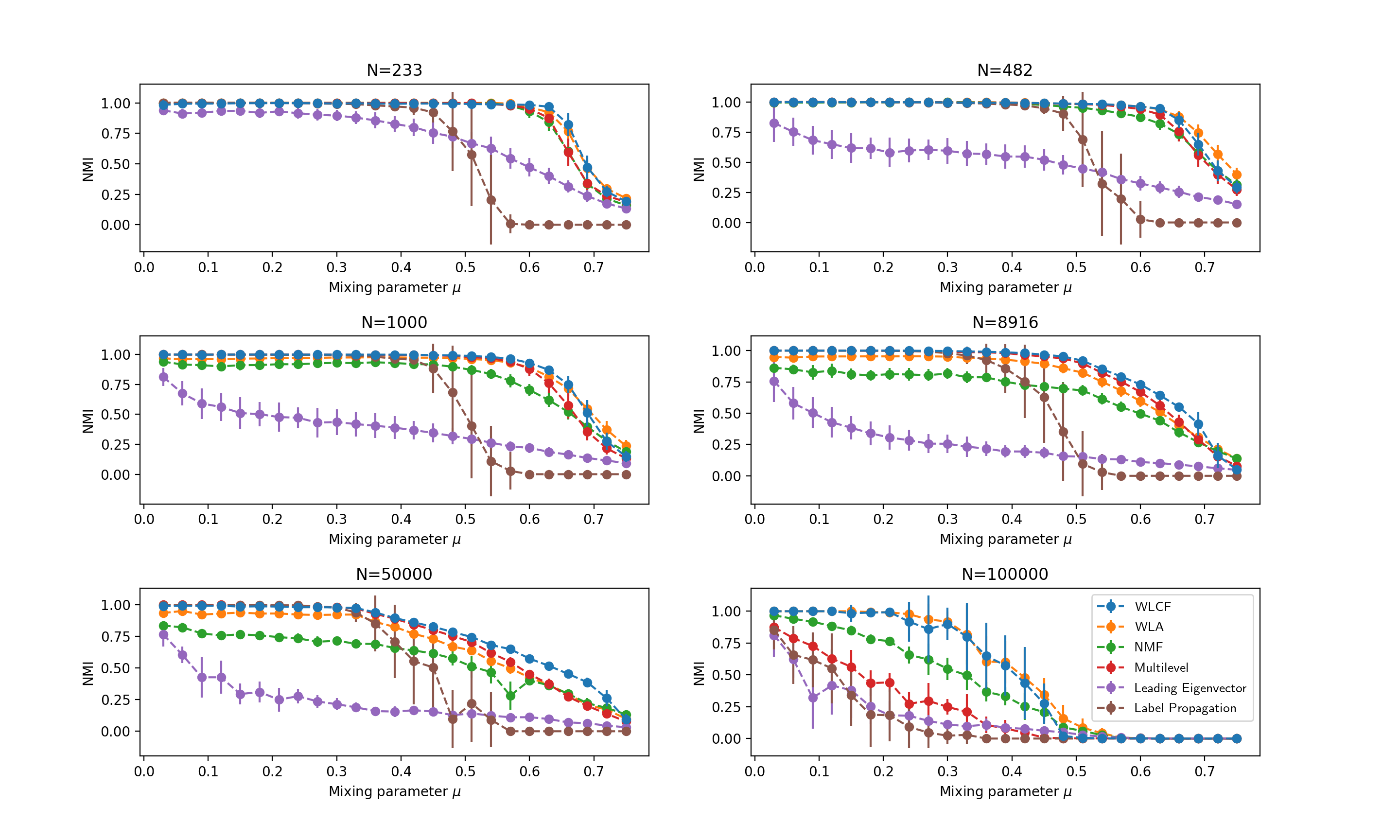}
\caption{
\textbf{Performance of WLCF and WLA on the LFR benchmark (NMI).}
In each panel, Normalized Mutual Information (NMI)~(Eq.~\eqref{eq:nmi}) is plotted as a function of
the mixing parameter $\mu$ for a given LFR network size $N$ (LFR network parameters are listed in Table~S1).
WLCF and WLA are compared with four state-of-the-art network community detection and clustering algorithms:
Multilevel~\cite{Blondel_2008}, Leading Eigenvector~\cite{Newman_2006}, Label Propagation~\cite{Raghavan_2007}, and Non-negative matrix factorization (NMF)~\cite{Lee1999,Lee2001}.
For each value of $N$ and $\mu$, we show $\langle \text{NMI} \rangle \pm \sigma_{\text{NMI}}$, where all averages and standard deviations are computed over independent network realizations.
}
\label{fig:NMI}
\end{figure}

Next, we have carried out an extensive comparison of the WLCF and WLA algorithms with four other state-of-the-art community network detection and clustering methods (Fig.~\ref{fig:NMI}).
Two methods, Multilevel~\cite{Blondel_2008} and Label Propagation~\cite{Raghavan_2007}, were chosen because they were recommended by the previous large-scale investigation of algorithm
performance on the LFR benchmark~\cite{Yang2016}. We also included Leading Eigenvector~\cite{Newman_2006} because its cluster bifurcation approach is similar to that employed
by WLCF (Fig.~\ref{fig:flow}). We used the network analysis package \texttt{igraph} (\texttt{https://igraph.org}) to implement Multilevel, Label Propagation, and Leading Eigenvector; all
parameters were set to their default values.

In addition, we used \texttt{scikit-learn} 
to implement the NMF clustering method,\footnote{\texttt{https://scikit-learn.org/stable/modules/generated/sklearn.decomposition.NMF.html}} with the coordinate descent solver (\textit{solver='cd'}),
Nonnegative Double Singular Value Decomposition (NNDSVD) initialization (\textit{init='nndsvd'})~\cite{nndsvd}, and all other parameters left at their default values.
Since NMF requires the number of clusters as input, we provided $m$, the exact number of communities in each LFR network. Both WLCF and WLA used $l_{max} = 8$. In WLCF,
random assignment of nodes to communities upon bifurcation was employed. Similar to NMF, WLA had to be provided with the exact number of communities $m$ as input.
Moreover, NMF-based clustering was used to initialize stand-alone WLA, since random partition of the network into $m$ communities in the beginning results in somewhat inferior performance,
as described below.

We observe that WLCF generally outperforms all other algorithms in terms of NMI,
with NMF and Multilevel being the most competitive alternatives. However, their performance tends to deteriorate faster for larger
networks. We also note that WLA provides a significant advantage over NMF (both algorithms require the exact number of clusters as input). As expected, the performance of all the algorithms degrades with $\mu$ since network communities become less well separated as $\mu$ increases.
Another measure of performance is the relative error in predicting the number of clusters, $\Delta_m = {|m^\star - m|}/{m}$, where $m^\star$ is the predicted and $m$ is the exact
number of communities in each LFR benchmark network. WLCF also outperforms Multilevel, Label Propagation, and Leading Eigenvector using this measure (Fig.~S2), especially 
with $\mu > 0.5$. The next best-performing algorithm is Multilevel, except for $N=10^5$ where Label Propagation performs much better than Multilevel but still worse than WCLF.
In summary, WLCF outperforms the other algorithms in terms of both NMI and $\Delta_m$ measures of prediction accuracy.

We have also explored how the performance of WLCF is affected by various hyperparameter, initialization and algorithmic choices within its main pipeline (Fig.~\ref{fig:flow}). 
In addition to the random assignment of nodes to two new communities at the bifurcation step which was used in the standard WLCF algorithm (Fig.~\ref{fig:NMI}), we have investigated the effects of more
sophisticated community initialization protocols that employ either NMF or NNDSVD-based node assignment to provide better initial conditions for WLA within the WLCF
pipeline (Fig.~S3).
However, the effect was found to be minor on the LFR benchmark, leading us to conclude that non-random community initialization is not necessary as part of the WLCF protocol.
Interestingly, there was a noticeable gain when stand-alone WLA was initialized with NMF-predicted rather than random communities (Fig.~S3). Apparently, gains related to NMF or NNSVD-based
WLA initialization largely disappear when the number of new communities is always two, as is the case in the WLCF bifurcation step. Another potential reason is the WLCF
community merge step, which may help rectify errors incurred by the randomly initialized WLA. 

Since WLA depends on the maximum number of random walk steps $l_{max}$, we have also investigated a version of WLCF in which WLA was run with $l_{max} = 8$ followed by $l_{max} = 1$ at every subsequent iteration of the main loop within WLA,
starting with $l_{max} = 8$. The alternation between high and low values of $l_{max}$ was designed to explore both large- and small-scale network structures; however, no substantial gain
was observed compared to WLA with $l_{max} = 8$ (Fig.~S3).
Finally, we have explored the overall role of WLA in the WLCF pipeline by replacing it completely with NMF-based node assignment (cf. purple curves in Fig.~S3).
Excluding WLA from the pipeline leads to significant degradation of the WLCF performance, leading us to conclude that the performance boost provided by WLA is indispensable
for the overall success of the WLCF algorithm.

We have also studied how the time complexity of WLCF and WLA scales with the network size $N$. We empirically observe power-law behavior of the runtime on the LFR networks from our dataset,
$T \sim N^\alpha$, with the scaling exponents ranging from 1.19 to 1.74 for WLCF and from 1.39 to 1.91 for WLA (Fig.~S4).
This relatively weak dependence on the network size leads us to conclude that both of our algorithms are capable of treating large-scale networks.



\subsection*{Real-World Networks}

\noindent \textbf{Eight networks.}
After exploring the performance of our algorithms on the LFR benchmark, we have applied WLCF to eight small- and medium-size real-world networks widely studied in the network
literature:  Bottlenose dolphins network~\cite{Dolphins}, Les Mis\'erables network~\cite{lesmis}, American college football teams network~\cite{Girvan_2002},
Jazz musicians network~\cite{jazz_musicians}, \textit{C. elegans} neural network~\cite{Celegans}, Erdos co-authorship network~\cite{Erdos02,NR},
Edinburgh associative thesaurus network~\cite{eatRS}, and High-energy theory (HET) citation network~\cite{Hepthenew} (see Supplementary Materials (SM) Methods for the details of each network).

\begin{table}[!htb]
\resizebox{\textwidth}{!}{%
\begin{tabular}{ |l|l|l|l|l|l|l|l|l|l|l|l| }
 \hline
 Network & $N$ & $\langle k \rangle$ & \multicolumn{2}{|c|}{WLCF} & \multicolumn{2}{|c|}{Multilevel} \\
  &  &  & \multicolumn{2}{|c|}{} & \multicolumn{2}{|c|}{} \\
 \hline
  & & & $\langle M \rangle \pm \sigma_M$ & $\langle N_\mathrm{cl} \rangle \pm \sigma_{N_\mathrm{cl}}$ & $\langle M \rangle \pm \sigma_M$ & $\langle N_\mathrm{cl} \rangle \pm \sigma_{N_\mathrm{cl}}$  \\
 \hline
 \hline
Dolphin groups & 62 & 5.13 & $0.5181 \pm 0.0123$ & $4.21 \pm 0.45$ & $0.5204 \pm 0.0029$ & $5.15 \pm 0.55$ \\
Les Mis\'erables characters & 77 & 6.60 & $0.5467 \pm 0.0109$ & $5.45 \pm 0.65$ & $0.5563 \pm 0.0028$ & $6.34 \pm 0.55$ \\
Football teams & 115 & 10.66 & $0.6023 \pm 0.0050$ & $9.75 \pm 0.54$ & $0.6039 \pm 0.0018$ & $9.69 \pm 0.52$ \\
Jazz musicians  & 198 & 27.70 & $0.4404 \pm 0.0034$ & $3.35 \pm 0.48$ & $0.4430 \pm 0.0025$ & $3.84 \pm 0.37$ \\
\textit{C. elegans} neurons & 297 & 15.80 & $0.3957 \pm 0.0086$ & $4.79 \pm 0.82$ & $0.4093 \pm 0.0054$ & $5.75 \pm 0.50$ \\
Erdos co-authors & 6927  & 3.42 & $0.6650 \pm 0.0097$ & $25.41 \pm 1.93$ & $0.6957 \pm 0.0018$ & $31.77 \pm 1.77$ \\
Thesaurus words & 23219 & 67.95 & $0.3201 \pm 0.0027$ & $7.62 \pm 0.81$ & $0.3149 \pm 0.0028$ & $12.20 \pm 1.12$ \\
HET citations & 27770 & 25.41 & $0.6529 \pm 0.0030$ & $16.37 \pm 1.11$ & $0.6554 \pm 0.0028$ & $171.56 \pm 1.83$ \\
 \hline
\end{tabular}
}
\caption{\textbf{Performance of community detection algorithms on real-world networks.} Shown are the average and the standard deviation of the modularity score $M$ (Eq.~\eqref{M:score}) and the number of clusters $N_\mathrm{cl}$ predicted by WLCF and Multilevel algorithms on 8 real-world networks (see SM Methods for network descriptions).
All statistics are computed using $10^2$ independent runs of each algorithm per network.
The networks are unweighted (i.e., all edge weights are set to $1.0$). $N$ is the number of nodes in the network and $\langle k \rangle$ is the average number of links per node, a measure of network sparseness.
}
\label{table:8net}
\end{table}

We find that WLCF and Multilevel produce comparable modularity scores (Table~\ref{table:8net}), while the performance of the Leading Eigenvector and the Label Propagation algorithms is worse overall (Table~S2). Interestingly, WLCF tends to predict fewer clusters than Multilevel, furnishing more interpretable partitions without a substantial loss in the modularity score.
To investigate the nature of the network partitions found by the four algorithms, we have also computed the distributions of internal edge density and cut ratio scores~\cite{Leskovec2010,Yang2013}
(Methods). Despite being normalized by the total number of possible links, both scores tend to correlate with the number of clusters into which the network is partitioned, since the internal edge density is high in small, densely connected clusters, whereas the cut ratio is low in large clusters with relatively few outside links.

We observe that WLCF clusters do not have the highest internal edge density scores: the scores tend to be consistently smaller than those of Multilevel clusters (Table~S3) and the results are mixed vs. Leading Eigenvector and Label Propagation clusters (Table~S4). The biggest discrepancies
can be traced to the differences in the number of clusters predicted by the four algorithms. For example, WLCF produces many fewer clusters in the HET citations network, resulting in much lower internal edge density scores. However, WLCF tends to produce lower cut ratio scores compared with the other three algorithms, a sign of more self-contained clusters with fewer external
links. Overall, we conclude that
WLCF optimizes modularity and cut ratio scores to a larger extent than internal edge density, partly because it partitions the network into fewer clusters.

We have also investigated how WLCF cluster predictions are affected by including edge weights. We have focused on two of the networks where edge weights are available in the primary data: Les Mis\'erables characters and Thesaurus words (see SM Methods for edge weight definitions). With the Les Mis\'erables characters network, we obtain
$\langle M \rangle \pm \sigma_M = 0.5621 \pm 0.0064$ and $\langle N_\mathrm{cl} \rangle \pm \sigma_{N_\mathrm{cl}} = 5.82 \pm 0.41$ over $10^2$ independent runs
of the WLCF algorithm when the weights are included. These results are similar to those on the unweighted network, and indeed $\langle \text{NMI} \rangle \pm \sigma_\text{NMI} = 0.78 \pm  0.05$
between weighted and unweighted network partitions, showing that they are fairly consistent. In contrast, for Thesaurus words we observe
$\langle M \rangle \pm \sigma_M = 0.4759 \pm 0.0069$ and $\langle N_\mathrm{cl} \rangle \pm \sigma_{N_\mathrm{cl}} = 15.30 \pm 1.62$, a much more modular network with twice as
many clusters compared to the unweighted version (Table~\ref{table:8net}). The low overlap between weighted and unweighted network clusters
($\langle \text{NMI} \rangle \pm \sigma_\text{NMI} = 0.30 \pm 0.02$) shows that the decision to include or disregard edge weights plays a major role in this case.
These findings underscore the necessity of the careful design of the experiments that generate primary data. \\



\begin{figure}
\centering
\includegraphics[scale=0.17]{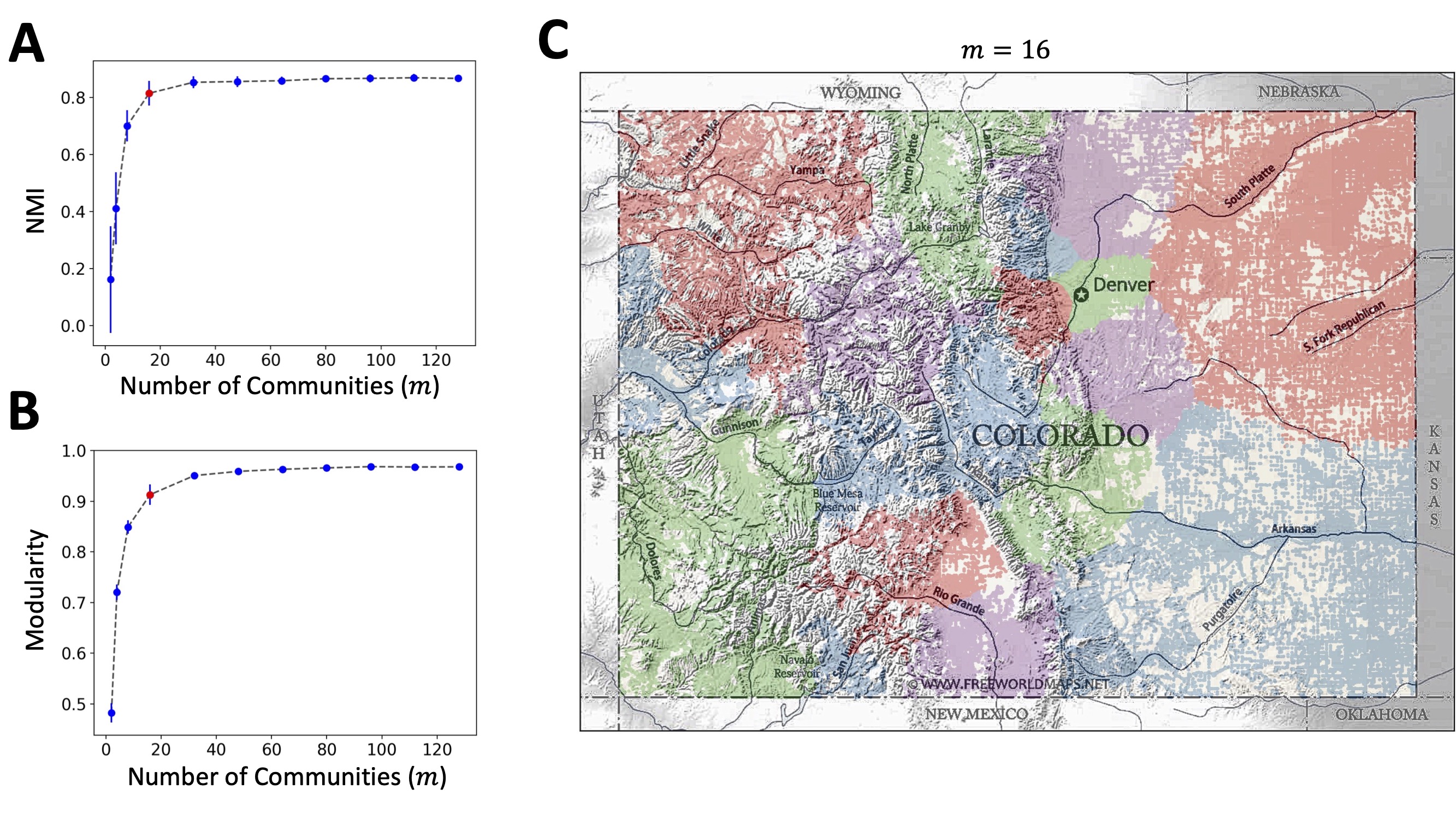}
\caption{
\textbf{WLA clustering of the Colorado road network.}
WLA was run 20 times for each value of $m$, with $m = \{ 2,4,8,16,32,48,64,80,96,112,128 \}$ (220 independent runs in total). Each run started from a random initial assignment
of nodes to communities and used $l_{\text{max}}=10^5$. Panel A: mean and standard deviation of the normalized mutual information (NMI) for the ensemble of all $\binom{20}{2}$ unique pairs
of network partitions for each value of $m$. Panel B: mean and standard deviation of the modularity score for 20 runs for each value of $m$. Panel C: visualization of one randomly chosen network partition
with $m = 16$ communities (shown as a red dot in panels A and B). Each node was assigned the color of its community and superimposed on a Colorado map using its longitude and latitude coordinates. The geographical map of the Colorado state was obtained from the website \texttt{freeworldmaps.net}
and rendered black-and-white. Colors were assigned to each community using the greedy coloring algorithm (\texttt{networkx.algorithms.coloring.greedy\_color}) from the NetworkX Python network analysis package (\texttt{https://networkx.org}). The coloring algorithm assigned 4 colors (red, green, blue and purple) to 16 communities such that no pair of adjacent communities have the same color.
}
\label{fig:COmap}
\end{figure}

\noindent \textbf{Colorado roadmap.} To investigate whether our approach can be applied to large-scale networks, we have chosen a graph defined
by geographical coordinates of road intersections and other landmarks in the state of Colorado.\footnote{\texttt{http://users.diag.uniroma1.it/challenge9/download.shtml}}
The network is very sparse, with $N = 435666$ nodes and ${\cal E} = 528533$ edges. We have made the network unweighted by assigning unit weights to each edge and run the WLA
algorithm on it multiple times (Fig.~\ref{fig:COmap}). We observe that with $m \le 16$, independent runs result in somewhat different cluster assignments, as can be seen from the lower NMI values and the error bars in Fig.~\ref{fig:COmap}A. However,
as the number of clusters increases, the assignment of nodes to clusters becomes more reproducible, with the NMI values around 0.87 and high consistency between the runs.
Similarly, the modularity score improves with the number of clusters, with the values around 0.97 for $m > 40$ (Fig.~\ref{fig:COmap}B).
These high values of modularity scores are not surprising since, given the sparseness of the network, it is relatively easy to partition the graph into smaller clusters that are only weakly connected to one another.

Fig.~\ref{fig:COmap}C shows a single randomly chosen realization of partitioning the network into $m=16$ clusters (Fig.~S5 contains three additional examples with $m=2,4,8$). In all of these examples, the results are intuitively compelling -- each cluster occupies a geographically contiguous region and the boundaries between neighboring communities often coincide with mountain ranges, major rivers, and other geographical landmarks. We conclude that our approach can be used to detect community structure in large-scale complex networks.

\section*{Discussion}

In this work, we have developed a novel approach to partitioning complex networks into non-overlapping communities. Networks that occur in nature and society often exhibit community structure, with
nodes within communities connected by more links than nodes in different communities (see e.g. Refs.~\cite{Girvan_2002,Radicchi2004}). However,
this structure is often challenging to detect and there may be many alternative solutions of similar quality, confronting community detection algorithms with a hard optimization problem.
The task of finding communities in networks is similar to a clustering problem in machine learning, in which, in the case of hard clustering, the dataset is divided into disjoint subsets
on the basis of pairwise distances between datapoints~\cite{Bishop:2006}. 

Our approach is based on the observation that short random walks that start in a given community will preferentially explore that community. 
To avoid potential issues related to finite sampling, we formally consider the limit of an infinite number of random walks which start from
all nodes in the network according to the connectivity of each node. For each random walk, the expected number of visits to each node in the network is computed exactly using the
transition matrix of the network. Since the total number of random walks is infinite, there is no sampling noise and the expected number of visits to each node provides an exact statistic,
which is used to assign nodes to communities in a Bayesian sense. The number of steps in each random walk, $l_{max}$, is a key hyperparameter of the algorithm: choosing
a very small value will mean that walks may not be able to reach some of the nodes within their own community, while choosing 
a very large value will make it more difficult to differentiate between communities (Fig.~S1). In other words, the value of $l_{max}$ determines the scale of the structures explored by the
diffusion process.

In practice, our algorithm, which we call the walk-likelihood algorithm, or WLA for short, is run iteratively starting from the initial condition that is either random or provided by another algorithm
such as non-negative matrix factorization (NMF)~\cite{Lee1999,Lee2001}. The algorithm is terminated once the partition of the network into $m$ communities
stops changing substantially from iteration to iteration. Since WLA requires the total number
of communities $m$ as input, we have created another algorithm, the walk-likelihood community finder, or WLCF, which uses WLA as a basic building block to produce the optimal number
of network communities $m_\text{opt}$ through global moves such as community bifurcation and merging (Fig.~\ref{fig:flow}).

Our main score for judging the success of the clustering procedure is the network modularity score~\cite{Newman_2004}, although we have also considered two additional measures:
the internal edge density and the cut ratio~\cite{Leskovec2010,Yang2013}. To benchmark WLA and WLCF against other algorithms in a controlled setting, we have employed the LFR benchmark which was created to provide a challenging test for community detection algorithms~\cite{Lancichinetti_2008}. On this benchmark, WLA and WLCF compare very favorably with several state-of-the-art
community detection and clustering algorithms (Figs.~\ref{fig:NMI},S2). 
Moreover, 
the dependence on the exact values of $l_{max}$ appears to be weak (Fig.~S3).

Another dataset we have considered consists of eight small- and medium-size real-world networks that are often investigated in the network science literature (Tables~\ref{table:8net},S2).
On this group of networks, WLCF produces modularity scores comparable to those predicted by another algorithm, Multilevel~\cite{Blondel_2008}, while partitioning the network into
fewer clusters. WLCF also tends to produce low cut ratio scores, a sign that it identifies self-contained clusters with few external links. However, WLCF clusters are not characterized by the highest
internal edge density scores compared to the other algorithms (Tables~S3,S4), probably because these scores increase trivially with the number of communities and WLCF tends to produce
fewer clusters.

Using a set of networks from the LFR benchmark, we find a power-law relation between the WLCF and WLA running times $T$ and the total number of nodes in the network: $T \sim N^\alpha$,
with the scaling exponent $1.0 < \alpha < 2.0$ that depends on the network type (Fig.~S4). Therefore, our approach can be used to analyze large-scale networks which may present
difficulties to other algorithms. To demonstrate this ability, we have applied WLA to a network of roads in the state of Colorado with almost half a million nodes (Figs.~\ref{fig:COmap},S5).
The results are geographically sensible, with neighboring clusters separated by major rivers, mountain ranges, or corresponding to urban agglomerations such as Denver
metropolitan area. 

To summarize, our computational framework for clustering and network community detection is efficient and robust with respect to the choice of initial conditions and hyperparameter values.
It compares favorably with several state-of-the-art algorithms. Although ideas centered on random walks and diffusion processes were previously explored in machine learning in the context
of diffusion maps~\cite{Coifman2005,Coifman2006,Delaporte2008} and spectral clustering~\cite{NMF3,Luxburg2007}, our approach is unique in its use of random walks to assign
nodes to communities probabilistically in a Bayesian sense. This is a significant extension of our previous work, which used conceptually similar ideas to infer properties of the entire
network, such as its size, on the basis of sparse exploration by random walks, but without partitioning the network into distinct communities~\cite{Willow}.
In the future, we will investigate both novel applications and algorithmic extensions of our approach, including
its adaptation to the soft clustering problem.   

\section*{Methods}

\noindent
\textbf{Network community metrics.}
Consider a network (undirected graph) with $N$ nodes, or vertices. The network is divided into $m$ non-overlapping communities, or clusters, with $N_c$ nodes in community $c = 1 \dots m$:
$N = \sum_{c=1}^m N_c$. The network contains ${\cal E}$ edges in total; we also define ${\cal I}_c$, the total number of internal edges that connect nodes within community $c$, and
${\cal E}_c$, the total number of external edges that connect nodes in community $c$ to nodes in all other communities.
Finally, a node $i$ $(i = 1 \dots N)$ has $k_i$ edges attached to it, such that ${\cal E} = (1/2) \sum_{i=1}^N k_i$ and $T_c = \sum_{i \in c} k_i$ is the total number of edge ends attached to the
nodes in community $c$.

With these definitions, the modularity score is given by~\cite{Newman_2004}:
\begin{equation} \label{M:score}
    M = \sum_{c=1}^m \left( e_{cc} - a_c^2 \right),
\end{equation}
where $e_{cc} = {\cal I}_c/{\cal E}$ is the fraction of all network edges that are internal to community $c$
and $a_c = T_c/2 {\cal E}$ is the fraction of all edge ends that are attached to the vertices in community $c$,
such that $a_c^2$ is the expected value of the fraction of edges internal to the community $c$ if the edges were placed at random.
Thus, the modularity score is a sum over differences between the observed and the expected fraction of internal edges in each community.
By construction, the positive modularity score indicates non-trivial groupings of nodes within the network with, on average, more connections between nodes
within each community than could be expected by chance.

We also introduce two additional metrics used to estimate the quality of network partitions into communities~\cite{Leskovec2010,Yang2013}:
(i) the internal edge density ${\cal D}_c = 2 {\cal I}_c/ N_c (N_c-1)$, which measures the fraction
of all possible internal edges observed in cluster $c$, averaged over all clusters: ${\cal D} = (1/m) \sum_{c=1}^m {\cal D}_c$ ;
(ii) the average cut ratio ${\cal R} = (1/m) \sum_{c=1}^m {\cal R}_c$, where ${\cal R}_c = {\cal E}_c/ N_c (N - N_c)$ is the fraction of all possible
external edges leaving the cluster.

\noindent
\textbf{Random walks on networks with communities.}
Consider a discrete-time random walk on an undirected network with weighted edges:
$\{w_{ij}\}$, where $w_{ij} = w_{ji}$ is the edge weight or rate of transmission from
node $i$ to $j$ (note that $w_{ij} = 1$ for unweighted networks). At each step the random walker jumps to its nearest
neighbor with probability $P(i \rightarrow j) = w_{ij}/w_{i}$, where
$w_{i}=\sum_{k \in nn(i)} w_{ik}$ is the connectivity of node $i$ and the sum is over all nearest neighbors of node $i$.
For unweighted networks, $w_{i}=k_{i}$, the total number of edges attached to node $i$.
We assume that the network has a community $c$ with $N_{c}$ nodes. Then the average return time (i.e., the average number of random walk steps) to a node $n \in c$, provided that
there are no transitions outside of the community, is given by $\W_{c}/w_n$~\cite{Noh2004,Condamin2007,Condamin2007b,Willow}, where
$\W_{c} = \sum_{i=1}^{N_c} w_{i}$ is the weighted size of all nodes in community $c$.
Note that for a set of nodes in community $c$, $S=\{n_{1}, n_{2},..., n_{N_{p}}\}$, the average return time to any of the nodes in set $S$
is given by $\W_{c} / \W_{p}$ in the absence of inter-community transitions, where
$\W_{p} = \sum_{i=1}^{N_p} w_{i}$ is the weighted size of all nodes in set $S$.

Assuming that the distribution of return times is exponential, or memoryless~\cite{Willow},
the probability to return to node $n \in c$ after exactly $\ell$ steps, with no transitions outside
of the community $c$, is given by
\begin{equation}
P(\ell)=\frac{w_n}{\W_{c}}e^{-(\sfrac{w_n}{\W_{c}})\ell}.
\end{equation}
It then follows that the probability of not making a return for $\ell$ steps (i.e., the survival probability) is
\begin{equation}
S(\ell)=e^{-(\sfrac{w_n}{\W_{c}})\ell}.
\end{equation}
Thus, the probability of making $\mathcal{K}$ returns to the node $n$ if $\ell_{c}$ steps are taken within the community $c$ is given by the Poisson distribution ${\cal P}$:

\begin{equation} \label{eq:pois}
P(\mathcal{K}|\ell_{c}) = \mathcal{P} \left(\mathcal{K},\frac{w_n\ell_{c}}{\W_{c}}\right) = \frac{1}{\mathcal{K}!}\left(\frac{w_n}{\W_{c}}\ell_{c}\right)^{\mathcal{K}}e^{-(w_{n}/\W_{c})\ell_{c}}.
\end{equation}
Accordingly, the mean number of visits to the node $n \in c$ is found to be
\begin{equation} \label{eq:mean}
E (\mathcal{K}|n) = \frac{w_n\ell_{c}}{\W_{c}}.
\end{equation}

Next, consider a stochastic process in which node $i$ is visited $\{ \kappa_{j} \}_{j=1}^p$ times after $p$ random walks
on a network with $m$ communities, labeled $c =1 \dots m$. Assume that during random walk $j$ ($j = 1 \dots p$), the random walker takes $\ell_{j c}$ steps on each community
$c$ of weighted size $\W_{c}$: $\{ \ell_{j c} \}_{j=1}^p$ (we adopt a convention that a jump from community $c'$ to $c$ is considered a step in community $c$).
Now, if the node $i \in c$, the probability of visiting this node $\{ \kappa_{j} \}_{j=1}^p$ times after $p$ random walks is given by

\begin{equation}
P(\{\kappa_{j}\} | i \in c, \{ \ell_{jc} \}_{j=1}^p) = \prod_{j=1}^{p} \mathcal{P} \left( \kappa_{j}, \frac{w_{i} \ell_{jc}}{\W_{c}} \right).
\end{equation}

If the community assignment of node $i$ is not known, we can use Bayes' theorem with uniform priors $P(i \in c) = m^{-1}$ to find the posterior probability that node $i \in c$:

\begin{equation} \label{eq:2}
P(i \in c | \{\kappa_{j}\}_{j=1}^p, \{\ell_{j c}\}_{j=1}^p) = \frac{1}{\mathcal{Z}} \prod_{j=1}^{p} \mathcal{P} \left( \kappa_{j}, \frac{w_{i} \ell_{jc}}{\W_{c}} \right),
\end{equation}
where the normalization constant $\mathcal{Z}$ is given by
\begin{equation}
\mathcal{Z} = \sum_{c=1}^m \left[ \prod_{j=1}^{p} \mathcal{P} \left( \kappa_{j}, \frac{w_{i} \ell_{j c}}{\W_{c}} \right) \right].
\end{equation}

\noindent
\textbf{Normalized mutual information.}
We use NMI~\cite{NMI} to quantify the similarity between network partitions $U$ and $U'$:
\begin{equation} \label{eq:nmi}
\text{NMI}(U,U')=\frac{ 2 \sum _{c=1}^{m} \sum _{c'=1}^{m'} P_{UU'}(c,c') \log {\left[ P_{UU'}(c,c')/(P_{U}(c) P_{U'}(c')) \right]} }{\sum_{c=1}^{m} P_{U}(c) \log P_{U}(c)
+ \sum_{c=1}^{m'} P_{U'}(c) \log P_{U'}(c)
},
\end{equation}
where $P_U(c) = N^{-1} \sum_{n=1}^N U_{nc}$, $P_{UU'}(c,c') = N^{-1} \sum_{n=1}^N U_{nc} U'_{nc'}$, and $m$ and $m'$ refer to the number of communities in the partitions $U$ and $U'$, respectively.
Note that NMI is always between $0$ and $1$, with $\text{NMI}(U,U')=1$ if and only if the partitions $U$ and $U'$
are exactly the same. Although Eq.~\eqref{eq:nmi} is valid for general values of $m$ and $m'$, we focus on $m=m'$ because WLA node reassignment procedure does not change
the number of communities.

\noindent
\textbf{Software availability.} A Python implementation of WLA and WLCF is available at \\
\texttt{https://github.com/lordareicgnon/Walk\_likelihood/}.

\clearpage




\section*{Acknowledgements}

AB and AVM were supported by a grant from the National Science Foundation (NSF MCB1920914).
AB would like to thank Abhishek Bhrushundi for suggesting the benchmarks to test our algorithms against and for many other discussions related to this project.

\section*{Supplementary Materials}
Supplementary Methods\\
Figs. S1 to S5\\
Tables S1 to S4\\




\newpage


\renewcommand{\thefigure}{S\arabic{figure}}
\setcounter{figure}{0}

\renewcommand{\thetable}{S\arabic{table}}
\setcounter{table}{0}

\section*{Supplementary Methods}

\subsection*{Additional details of eight real-world networks}

\begin{itemize}

    \item \textbf{Bottlenose dolphins network:} A network of a group of dolphins from Doubtful Sound, New Zealand observed by David Lusseau, a researcher at the University of Aberdeen~\cite{Dolphins}. Every time a school of dolphins was encountered, each dolphin in the group was identified using natural markings on the dorsal fin. This information was utilized to form a social network where each node represents a dolphin and edges represent their preferred companionship.
    
    \item \textbf{Les Mis\'erables network:} A network of co-appearances of the characters in the novel Les Mis\'erables by Victor Hugo~\cite{lesmis}. Each node represents a character and each edge represents their co-occurrence in the novel's chapters. Edge weights are the number of chapters in which the two characters have appeared together. 
    
    \item \textbf{American college football teams network:} A network of all Division I college football games during the regular season in Fall 2000, with each node indicating a college team and the edge weight indicating the number of games between teams~\cite{Girvan_2002}.
    
    \item \textbf{Jazz musicians network:} This is a network of collaborations between jazz musicians~\cite{jazz_musicians}. Each node corresponds to a jazz musician and an edge denotes that two musicians have played together in a band.
    
    \item \textbf{\textit{C. elegans} neural network:} Each node in the network represents a neuron and each edge represents the neuron's connection with other neurons~\cite{Celegans}. Edge directionality was removed from the graph following Watts and Strogatz~\cite{Watts1998}.
    
    \item \textbf{Erdos co-authorship network:} A network which includes Paul Erdos, his co-authors, and their co-authors. Each node represents an author and there is an edge between two authors if they have co-authored a paper~\cite{Erdos02,NR}.
    
    \item \textbf{Edinburgh associative thesaurus network:} A network of word associations based on the word association counts collected from British university students around 1970.
Nodes are English words and a link between A and B denotes that the word B was given as a response to the stimulus word A~\cite{eatRS}. Edge weights are the number of times B was given in response to A. The graph was made non-directional by symmetrizing the edge weights.
    
    \item \textbf{High-energy theory citation network:} A citation network of high-energy physics theorists. Each node represents an author and there is an edge between two authors if they have cited each other in their papers~\cite{Hepthenew}.
    
\end{itemize}

\newpage
\section*{Supplementary Figures}

\begin{figure}[!htb]
\centering
\includegraphics[scale=0.4]{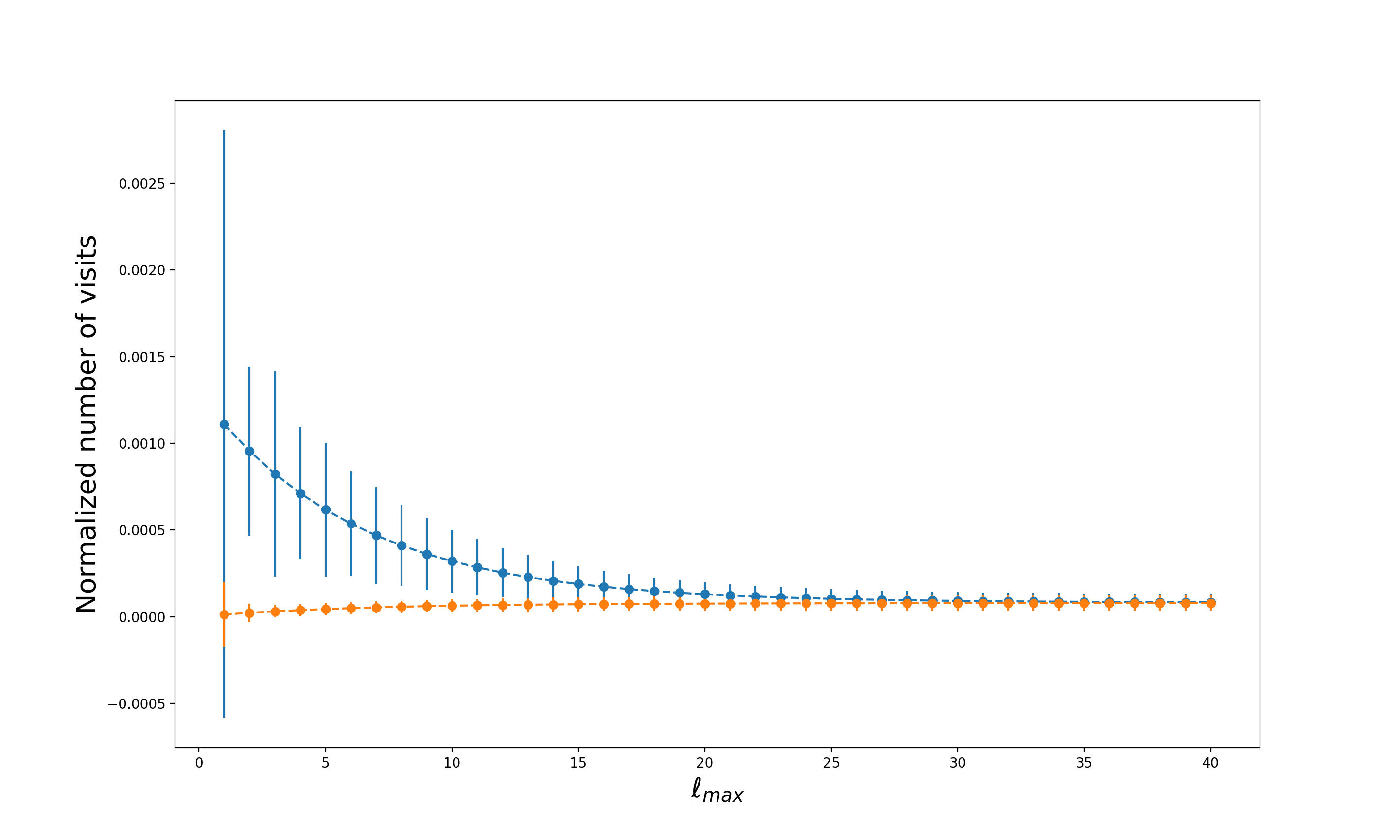}
\caption{
\textbf{The effect of $l_{max}$ on network exploration.}
A single realization of the LFR network with $\mu=0.15$ and $N=1000$ was used to plot the mean and the standard deviation of the normalized number of visits averaged over all pairs of nodes
that belong to the same community ($K_1$, blue curve), and different communities ($K_2$, orange curve). Specifically, we compute $\kappa_{ij}^{(l_{max})}/w_j = (\sum_{l=1}^{l_{max}}  A^{l})_{ij}$,
where $\kappa_{ij}^{(l_{max})}$ is the number of visits to node $j$ for the ensemble of random walks that start from node $i$ and make $l_{max}$ steps, and $w_j$ is the connectivity of node $j$.
Then $\langle K_1 \rangle = \langle \kappa_{ij}^{(l_{max})}/w_j \rangle_{i,j \in \text{same community}}$ and $\langle K_2 \rangle = \langle \kappa_{ij}^{(l_{max})}/w_j \rangle_{i,j \in \text{different communities}}$,
with standard deviations for both quantities computed using the same sets of node pairs.
}
\label{fig:lmax}
\end{figure}

\begin{figure}[!htb]
\centering
\hspace*{-1.5cm}
\includegraphics[scale=0.55]{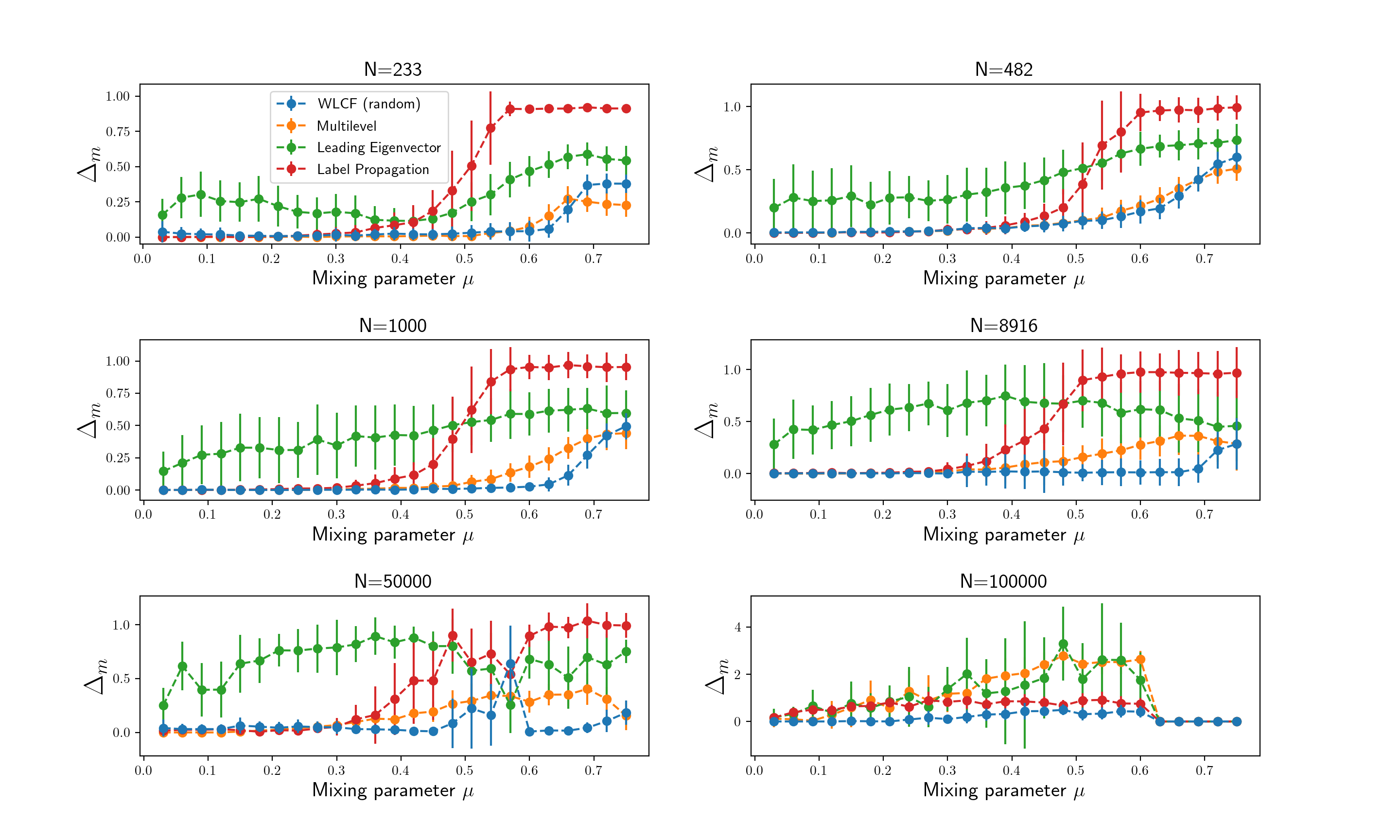}
\caption{
\textbf{Performance of WLCF and WLA on the LFR benchmark (number of communities).}
In each panel, relative deviation $\Delta_m$ between the predicted number of communities $m^\star$ and the exact number of communities $m$,
$\Delta_m = {|m^\star - m|}/{m}$,
is plotted as a function of the mixing parameter $\mu$ for a given LFR network size $N$ (LFR network parameters are listed in Table~\ref{table:LFR}).
WLCF and WLA are compared with four state-of-the-art network community detection and clustering algorithms:
Multilevel~\cite{Blondel_2008}, Leading Eigenvector~\cite{Newman_2006}, Label Propagation~\cite{Raghavan_2007}, and Non-negative matrix factorization (NMF)~\cite{Lee1999,Lee2001}.
For each value of $N$ and $\mu$, we show $\langle \Delta_m \rangle \pm \sigma_{\Delta_m}$, where all averages and standard deviations are computed over independent network realizations.
}
\label{fig:num_comm}
\end{figure}

\begin{figure}
\centering
\hspace*{-0.5cm}
\includegraphics[scale=0.22]{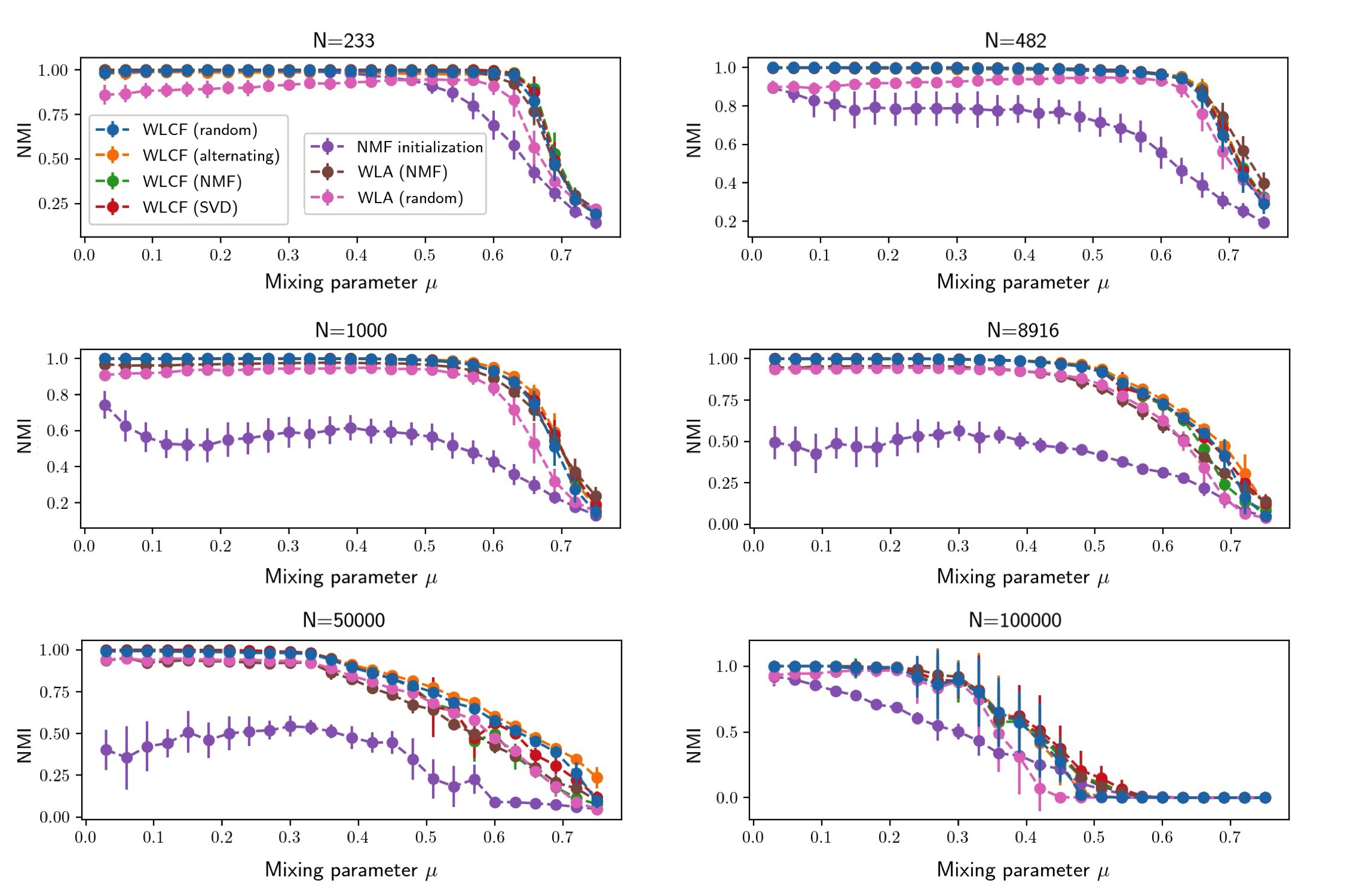}
\caption{
\textbf{Performance of different versions of WLCF and WLA on the LFR benchmark (NMI).}
In each panel, Normalized Mutual Information (NMI) 
is plotted as a function of the mixing parameter $\mu$ for a given LFR network size $N$ (LFR network parameters are listed in Table~\ref{table:LFR}).
The WLCF versions are: WLCF(random), same as WLCF in Fig.~2; WLCF(alternating), same as WLCF(random) but with $l_{max} = 8$ and $l_{max} = 1$ alternating
at each subsequent step within WLA, starting from $l_{max} = 8$; WLCF(NMF), same as WLCF(random) but with two communities determined by NMF~\cite{Lee1999,Lee2001}
rather than created randomly at the community bifurcation step, before applying WLA; WLCF(SVD), same as WLCF(random) but with two communities determined by NNSVD~\cite{nndsvd}
rather than created randomly at the community bifurcation step, before applying WLA; NMF initialization, same as WLCF(NMF) but without the WLA step, such that
the node community identities are determined solely by NMF. The WLA versions are: WLA(NMF), same as WLA in Fig.~2; WLA(random), same as WLA(NMF) but with all nodes split into
$m$ communities randomly in the beginning rather than assigned by NMF.
For each value of $N$ and $\mu$, we show $\langle \text{NMI} \rangle \pm \sigma_{\text{NMI}}$, where all averages and standard deviations are computed over independent network realizations.
}
\label{fig:NMIe}
\end{figure}

\begin{figure}[!htb]
\centering
\includegraphics[scale=0.18]{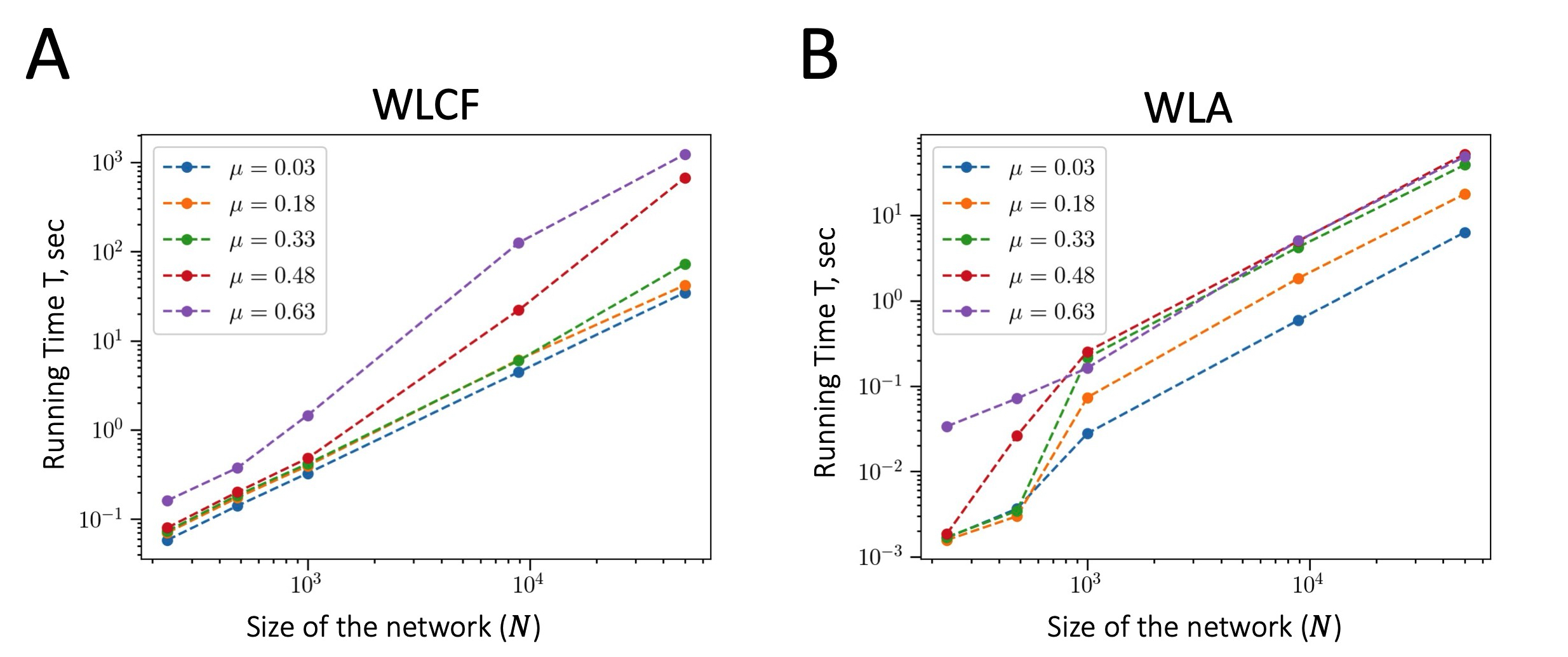}
\caption{
\textbf{Runtime scaling of WLCF and WLA.}
Shown are the wall times $T$ (in seconds) for WLCF (A) and WLA (B) applied to the LFR benchmark networks with different mixing parameters $\mu$, as a function of the network size $N$.
The wall times are averaged over all LFR network realizations with the same $N$ and $\mu$. Each curve is fitted to the power-law expression:
$T \sim N^\alpha$, yielding scaling exponents $\alpha = \left\{ 1.19, 1.20, 1.27, 1.69, 1.74 \right\}$ for WLCF (A) and
$\alpha = \left\{ 1.56, 1.79, 1.91, 1.79, 1.39 \right\}$ for WLA (B), for $\mu = \left\{ 0.03, 0.18, 0.33, 0.48, 0.63 \right\}$, respectively.
WLA was supplied with the exact number of communities $m$ for every network realization.
}
\label{fig:time}
\end{figure}

\begin{figure}[!htb]
\centering
\includegraphics[scale=0.20]{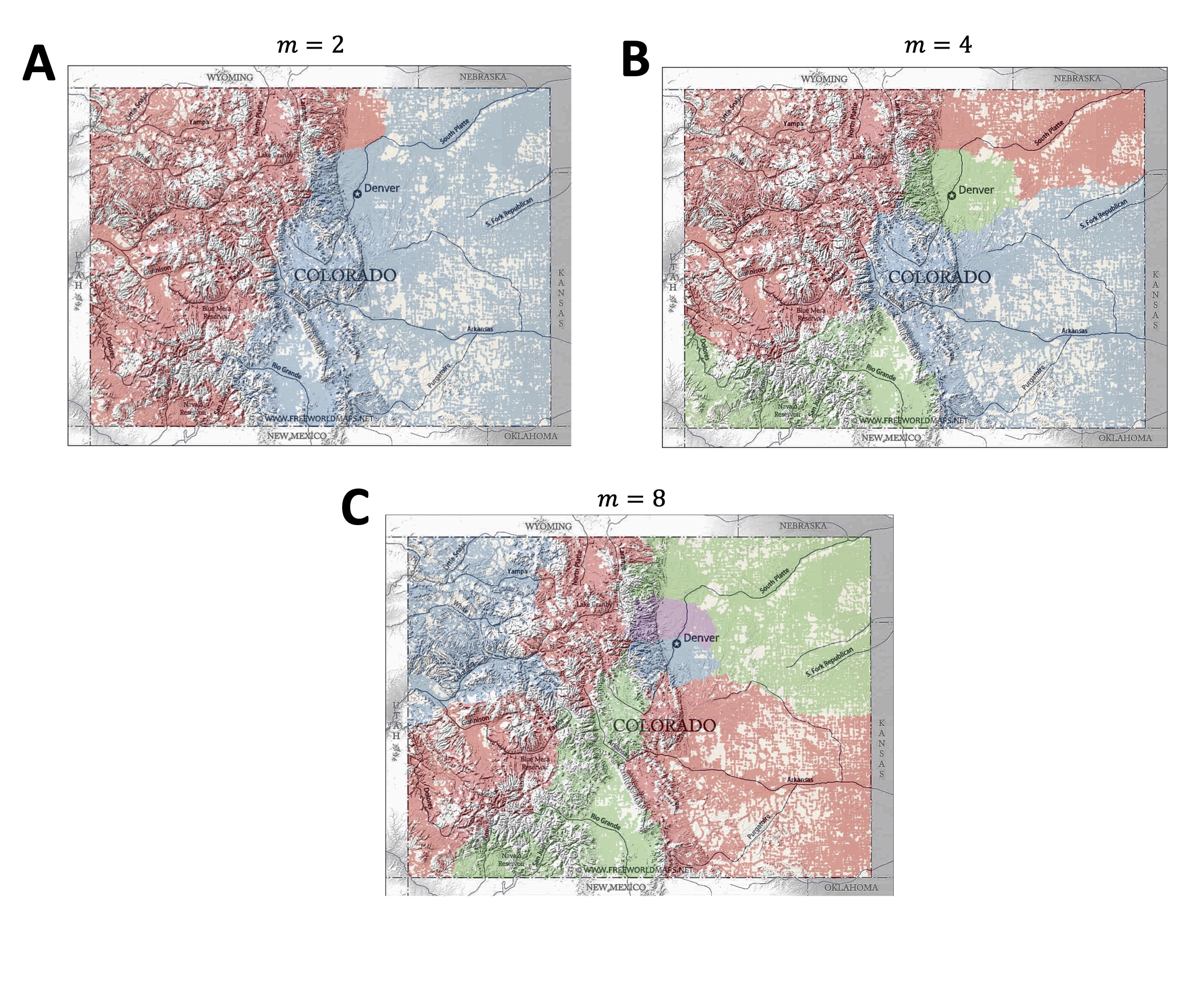}
\caption{
\textbf{WLA clustering of the Colorado road network: representative partitions.}
Shown are WLA partitions of the Colorado road network into $m=$ 2 (A), 4 (B), and 8 (C) communities. For each $m$, a single partition was randomly chosen from the 20 independent runs
described in the Fig.~3 caption. The community coloring scheme used 2 (A), 3 (B), and 4 (C) distinct colors; all other details of the color assignment are as in the Fig.~3 caption.
}
\label{fig:COmap:extra}
\end{figure}

\clearpage

\newpage
\section*{Supplementary Tables}

\begin{table}[!htb]
\begin{tabular}{ |l|l|  }
 \hline
 \textbf{Parameter} & \textbf{Value}\\
 \hline
 \hline
 Number of nodes $N$ & (233, 482, 1000, 8916, 50000, 100000)\\
 \hline
Maximum degree & $0.1N$\\
\hline
Maximum community size & $0.1N$\\
\hline
Average degree & $20$\\
\hline
Community size distribution exponent $\beta$ & $-1$\\
\hline
Degree distribution exponent $\gamma$ & $-2$\\
\hline
Mixing coefficient $\mu$ & $(0.03, 0.06, \dots, 0.75)$\\
\hline
\end{tabular}
\caption{
\textbf{Parameters of the networks in the LFR benchmark}~\cite{Lancichinetti_2008}.
}
\label{table:LFR}
\end{table}

\begin{table}[!htb]
\resizebox{\textwidth}{!}{%
\begin{tabular}{ |l|l|l|l|l|l|l|l| }
 \hline
 Network & $N$ & $\langle k \rangle$ & \multicolumn{2}{|c|}{Leading} & \multicolumn{2}{|c|}{Label} \\
  &  &  & \multicolumn{2}{|c|}{Eigenvector} & \multicolumn{2}{|c|}{Propagation} \\
 \hline
  & & & $M$ & $N_\mathrm{cl}$ & $\langle M \rangle \pm \sigma_M$ & $\langle N_\mathrm{cl} \rangle \pm \sigma_{N_\mathrm{cl}}$ \\
 \hline
 \hline
Dolphin groups &  62  & 5.13 & 0.4912 & 5 & $0.4722 \pm 0.0591$ & $3.72 \pm 0.94$ \\
Les Mis\'erables characters & 77 & 6.60 & 0.5323 & 8 & $0.5016 \pm 0.0672$ & $5.43 \pm 1.06$ \\
Football teams & 115 & 10.66 & 0.4926 & 8 & $0.5899 \pm 0.0141$ & $10.61 \pm 1.14$ \\
Jazz musicians & 198 & 27.70 & 0.3936 & 3 & $0.3472 \pm 0.0975$ & $2.76 \pm 0.71$ \\
\textit{C. elegans} neurons & 297  & 15.80 & 0.3415 & 5 & $0.0763 \pm 0.1059$ & $1.35 \pm 0.48$ \\
Erdos co-authors & 6927  & 3.42 & 0.5979 & 27 & $0.5940 \pm 0.0096$ & $319.88 \pm 26.39$ \\
Thesaurus words & 23219  & 67.95 & 0.2577 & 4 & $0 \pm 0$ & $1 \pm 0$ \\
HET citations & 27770 & 25.41 & 0.5010 & 152 & $0.3554 \pm 0.1046$ & $497.86 \pm 38.42$ \\
 \hline
\end{tabular}
}
\caption{\textbf{Performance of community detection algorithms on real-world networks.} Same as Table~1 in the main text but for the Leading Eigenvector and Label Propagation algorithms.
Since Leading Eigenvector is not stochastic, $M$ and $N_\mathrm{cl}$ resulting from a single run are reported for each network.
For Label Propagation, the statistics are computed using $10^2$ independent runs per network.
}
\label{table:8net:part2}
\end{table}

\begin{table}[!htb]
\resizebox{\textwidth}{!}{%
\begin{tabular}{ |l|l|l|l|l|l|l|l|l|l| }
 \hline
 Network & $N$ & $\langle k \rangle$ & \multicolumn{2}{|c|}{WLCF} & \multicolumn{2}{|c|}{Multilevel} \\
  &  &  & \multicolumn{2}{|c|}{} & \multicolumn{2}{|c|}{} \\
 \hline
  & & & $\langle \mathcal{D} \rangle \pm \sigma_\mathcal{D}$ & $\langle \mathcal{R} \rangle \pm \sigma_\mathcal{R}$ & $\langle \mathcal{D} \rangle \pm \sigma_\mathcal{D}$ & $\langle \mathcal{R} \rangle \pm \sigma_\mathcal{R}$ \\
 \hline
 \hline
Dolphin groups & 62 & 5.13 & $0.3364 \pm 0.0352$ & $0.1468 \pm 0.0507$ & $0.3593 \pm 0.0292$ & $0.2054 \pm 0.0439$ \\
Les Mis\'erables & 77 & 6.60 & $0.4163 \pm 0.0402$ & $0.1258 \pm 0.0260$ & $0.4605 \pm 0.0427$ & $0.1568 \pm 0.0210$ \\
characters &  &  &  &  &  &  \\
Football teams & 115 & 10.66 & $0.7420 \pm 0.0216$ & $0.3069 \pm 0.0117$ & $0.7504 \pm 0.0216$ & $0.3145 \pm 0.0159$ \\
Jazz musicians & 198 & 27.70 & $0.3646 \pm 0.0397$ & $0.1925 \pm 0.1543$ & $0.4034 \pm 0.0326$ & $0.3886 \pm 0.4245$ \\
\textit{C. elegans} & 297 & 15.80 & $0.1732 \pm 0.0261$ & $0.1296 \pm 0.0476$ & $0.1989 \pm 0.0207$ & $0.1772 \pm 0.0574$ \\
neurons &  &  &  &  &  &  \\
Erdos co-authors & 6927  & 3.42 & $0.0161 \pm 0.0047$ & $0.0045 \pm 0.0008$ & $0.0202 \pm 0.0028$ & $0.0044 \pm 0.0005$ \\
Thesaurus words & 23219 & 67.95 & $0.0042 \pm 0.0003$ & $0.0050 \pm 0.0008$ & $0.0077 \pm 0.0031$ & $0.0139 \pm 0.0069$ \\
HET citations & 27770 & 25.41 & $0.0164 \pm 0.0037$ & $0.0051 \pm 0.0009$ & $0.7703 \pm 0.0055$ & $0.0027 \pm 0.0008$ \\
 \hline
\end{tabular}
}
\caption{\textbf{Performance of community detection algorithms on real-world networks.} Shown are the average and the standard deviation of the internal edge density $\mathcal{D}$
and the cut ratio $\mathcal{R}$ averaged over all clusters that were predicted by WLCF and Multilevel algorithms on 8 real-world networks (see Supplementary Methods for the details of the networks).
All statistics are computed using the results of $10^2$ independent runs of each algorithm on each network (same runs as in Table~1).
The networks are unweighted (i.e., all edge weights are set to $1.0$). $N$ is the number of nodes in the network and $\langle k \rangle$ is the average number of links per node,
a measure of network sparseness.
}
\label{table:8net:RD}
\end{table}

\begin{table}[!htb]
\resizebox{\textwidth}{!}{%
\begin{tabular}{ |l|l|l|l|l|l|l|l| }
 \hline
 Network & $N$ & $\langle k \rangle$ & \multicolumn{2}{|c|}{Leading} & \multicolumn{2}{|c|}{Label} \\
  &  &  & \multicolumn{2}{|c|}{Eigenvector} & \multicolumn{2}{|c|}{Propagation} \\
 \hline
  & & & $\mathcal{D}$ & $\mathcal{R}$ & $\langle \mathcal{D} \rangle \pm \sigma_\mathcal{D}$ & $\langle \mathcal{R} \rangle \pm \sigma_\mathcal{R}$ \\
 \hline
 \hline
Dolphin groups &  62  & 5.13 & 0.3319 & 0.1645 & $0.3286 \pm 0.0853$ & $0.1688 \pm 0.1564$ \\
Les Mis\'erables characters & 77 & 6.60 & 0.3835 & 0.7361 & $0.4972 \pm 0.0735$ & $0.1649 \pm 0.1073$ \\
Football teams & 115 & 10.66 & 0.5833 & 0.3890 & $0.8050 \pm 0.0434$ & $0.4407 \pm 0.0913$ \\
Jazz musicians & 198 & 27.70 & 0.3240 & 0.1180 & $0.3563 \pm 0.1098$ & $0.1223 \pm 0.1494$ \\
\textit{C. elegans} neurons & 297 & 15.80 & 0.1506 & 0.1327 & $0.0778 \pm 0.0339$ & $0.0129 \pm 0.0180$ \\
Erdos co-authors & 6927  & 3.42 & 0.0688 & 0.1512 & $0.1777 \pm 0.0044$ & $0.1206 \pm 0.0134$ \\
Thesaurus words & 23219  & 67.95 & 0.0026 & 0.0026 & $0.0012 \pm 0$ & -- \\
HET citations & 27770 & 25.41 & 0.8436 & 0.2507 & $0.6057 \pm 0.0183$ & $0.1400 \pm 0.0109$ \\
 \hline
\end{tabular}
}
\caption{\textbf{Performance of community detection algorithms on real-world networks.} Same as Table~\ref{table:8net:RD} but for the Leading Eigenvector and Label Propagation algorithms.
Since Leading Eigenvector is not stochastic, $\mathcal{D}$ and $\mathcal{R}$ are based on the single run reported for each network in Table~\ref{table:8net:part2}.
For Label Propagation, the statistics are computed using $10^2$ independent runs of each algorithm on each network (same runs as in Table~\ref{table:8net:part2}).
}
\label{table:8net:RD:part2}
\end{table}

\clearpage



\begin{thebibliography}{10}

\bibitem{Albert2002}
R.~Albert, A.~L. Barab\'{a}si, {\it Rev Mod Phys\/} {\bf 74}, 47 (2002).

\bibitem{Barrat2008}
A.~Barrat, M.~Barthelemy, A.~Vespignani, {\it Dynamical Processes on Complex
  Networks\/} (Cambridge University Press, Cambridge, UK, 2008).

\bibitem{Newman2010}
M.~E.~J. Newman, {\it Networks: {A}n Introduction\/} (Oxford University Press,
  Oxford, UK, 2010).

\bibitem{Girvan_2002}
M.~Girvan, M.~E.~J. Newman, {\it Proc Nat Acad Sci USA\/} {\bf 99}, 7821
  (2002).

\bibitem{Radicchi2004}
F.~Radicchi, C.~Castellano, F.~Cecconi, V.~Loreto, D.~Parisi, {\it Proc Nat
  Acad Sci USA\/} {\bf 101}, 2658 (2004).

\bibitem{Reichardt2004}
J.~Reichardt, S.~Bornholdt, {\it Phys Rev Lett\/} {\bf 93}, 218701 (2004).

\bibitem{Newman_2004}
M.~E.~J. Newman, {\it Phys Rev E\/} {\bf 69}, 066133 (2004).

\bibitem{Newman_2004b}
M.~E.~J. Newman, {\it Eur Phys J B\/} {\bf 38}, 321 (2004).

\bibitem{Palla2005}
G.~Palla, I.~Derenyi, I.~Farkas, T.~Vicsek, {\it Nature\/} {\bf 435}, 814
  (2005).

\bibitem{Walktrap2005}
P.~Pons, M.~Latapy, {\it Computer and Information Sciences - ISCIS 2005\/},
  P.~Yolum, T.~G{\"u}ng{\"o}r, F.~G{\"u}rgen, C.~{\"O}zturan, eds.
  (Springer-Verlag, Berlin, Germany, 2005), pp. 284--293.

\bibitem{Newman_2006}
M.~E.~J. Newman, {\it Phys Rev E\/} {\bf 74}, 036104 (2006).

\bibitem{Raghavan_2007}
U.~N. Raghavan, R.~Albert, S.~Kumara, {\it Phys Rev E\/} {\bf 76}, 036106
  (2007).

\bibitem{Blondel_2008}
V.~D. Blondel, J.-L. Guillaume, R.~Lambiotte, E.~Lefebvre, {\it J Stat Mech:
  Theory and Experiment\/} {\bf 2008}, P10008 (2008).

\bibitem{Gasch2000}
A.~P. Gasch, {\it et~al.\/}, {\it Mol. Biol. Cell\/} {\bf 11}, 4241 (2000).

\bibitem{Mihalik2011}
A.~Mihalik, P.~Csermely, {\it PLOS Comp Biol\/} {\bf 7}, e1002187 (2011).

\bibitem{Clauset_2004}
A.~Clauset, M.~E.~J. Newman, C.~Moore, {\it Phys Rev E\/} {\bf 70}, 066111
  (2004).

\bibitem{Rosvall2007}
M.~Rosvall, C.~T. Bergstrom, {\it Proc Nat Acad Sci USA\/} {\bf 104}, 7327
  (2007).

\bibitem{Yang2016}
Z.~Yang, R.~Algesheimer, C.~J. Tessone, {\it Sci Rep\/} {\bf 6}, 30750 (2016).

\bibitem{Leskovec2010}
J.~Leskovec, K.~Lang, M.~Mahoney, {\it Proceedings of the 19th International
  Conference on World Wide Web, WWW 2010\/} (2010), pp. 1--10.

\bibitem{Yang2013}
Y.~Yang, Y.~Sun, S.~Pandit, N.~V. Chawla, J.~Han, {\it Perspective on
  Measurement Metrics for Community Detection Algorithms\/} (Springer,
  Dordrecht, The Netherlands, 2013), pp. 227--242.

\bibitem{Bishop:2006}
C.~M. Bishop, {\it {Pattern Recognition and Machine Learning}\/} (Springer, New
  York, NY, 2006).

\bibitem{Lee1999}
D.~D. Lee, H.~S. Seung, {\it Nature\/} {\bf 401}, 788 (1999).

\bibitem{Lee2001}
D.~Lee, H.~S. Seung, {\it Advances in Neural Information Processing Systems\/},
  T.~Leen, T.~Dietterich, V.~Tresp, eds. (MIT Press, 2001), vol.~13, pp. 1--7.

\bibitem{nndsvd}
C.~Boutsidis, E.~Gallopoulos, {\it Pattern Recognition\/} {\bf 41}, 1350
  (2008).

\bibitem{NMF4}
D.~Kuang, C.~Ding, H.~Park, {\it Proceedings of the 2012 SIAM International
  Conference on Data Mining (SDM)\/} (2012), pp. 106--117.

\bibitem{NMF3}
C.~Ding, X.~He, H.~D. Simon, {\it Proceedings of the 2005 SIAM International
  Conference on Data Mining (SDM)\/} (2005), pp. 606--610.

\bibitem{Luxburg2007}
U.~{von Luxburg}, {\it Stat Comput\/} {\bf 17}, 395 (2007).

\bibitem{Coifman2005}
R.~R. Coifman, {\it et~al.\/}, {\it Proc Nat Acad Sci USA\/} {\bf 102}, 7426
  (2005).

\bibitem{Coifman2006}
R.~R. Coifman, S.~Lafon, {\it Appl Comput Harmon Anal\/} {\bf 21}, 5 (2006).

\bibitem{Delaporte2008}
J.~{de la Porte}, B.~M. Herbst, W.~A. Hereman, S.~J. {van der Walt}, An
  introduction to diffusion maps (2008). {\it unpublished}.

\bibitem{Willow}
W.~B. Kion-Crosby, A.~V. Morozov, {\it Phys Rev Lett\/} {\bf 121}, 038301
  (2018).

\bibitem{NMI}
A.~Strehl, J.~Ghosh, {\it J Mach Learn Res\/} {\bf 3}, 583 (2002).

\bibitem{Lancichinetti_2008}
A.~Lancichinetti, S.~Fortunato, F.~Radicchi, {\it Phys Rev E\/} {\bf 78},
  046110 (2008).

\bibitem{Dolphins}
D.~Lusseau, {\it Proc Biol Sci\/} {\bf 270}, S186 (2003).

\bibitem{lesmis}
D.~E. Knuth, {\it The Stanford GraphBase -- a platform for combinatorial
  computing\/} (ACM Press, New York, NY, 1993).

\bibitem{jazz_musicians}
P.~M. Gleiser, L.~Danon, {\it Adv Compl Syst\/} {\bf 6}, 565 (2003).

\bibitem{Celegans}
J.~G. White, E.~Southgate, J.~N. Thomson, S.~Brenner, {\it Philos Trans R Soc
  Lond Ser B Biol Sci\/} {\bf 314}, 1 (1986).

\bibitem{Erdos02}
J.~Kunegis, {\it Proc. Int. Conf. on World Wide Web Companion\/} (2013), pp.
  1343--1350.

\bibitem{NR}
R.~A. Rossi, N.~K. Ahmed, {\it Proceedings of the Twenty-Ninth AAAI Conference
  on Artificial Intelligence\/} (2015), pp. 4292--4293.

\bibitem{eatRS}
G.~R. Kiss, C.~Armstrong, R.~Milroy, J.~Piper, {\it The computer and literary
  studies\/}, A.~J. Aitkin, R.~W. Bailey, N.~Hamilton-Smith, eds. (University
  Press, Edinburgh, UK, 1973).

\bibitem{Hepthenew}
J.~Gehrke, P.~Ginsparg, J.~Kleinberg, {\it ACM SIGKDD Explorations
  Newsletter\/} {\bf 5}, 149 (2003).

\bibitem{Noh2004}
J.~D. Noh, H.~Rieger, {\it Phys Rev Lett\/} {\bf 92}, 118701 (2004).

\bibitem{Condamin2007}
S.~Condamin, O.~B{\'{e}}nichou, V.~Tejedor, R.~Voituriez, J.~Klafter, {\it
  Nature\/} {\bf 450}, 77 (2007).

\bibitem{Condamin2007b}
S.~Condamin, O.~B{\'{e}}nichou, M.~Moreau, {\it Phys Rev E\/} {\bf 75}, 021111
  (2007).

\bibitem{Watts1998}
D.~J. Watts, S.~H. Strogatz, {\it Nature\/} {\bf 393}, 440 (1998).

\end{thebibliography}

\end{document}